\documentclass[prd,preprint,superscriptaddress,
  tightenlines,nofootinbib, eqsecnum,showpacs]{revtex4}

\usepackage{amsmath}
\usepackage{amsfonts}
\usepackage{amssymb}
\usepackage{bm}
\usepackage{hyperref}
\usepackage{mathrsfs}
\usepackage{graphicx}

\usepackage{empheq}

\usepackage{ulem}
\usepackage[usenames]{color}

\definecolor{darkgreen}{rgb}{0,0.5,0}

\hypersetup{
    bookmarks=true,         
    unicode=false,          
    pdftoolbar=true,        
    pdfmenubar=true,        
    pdffitwindow=false,     
    pdfstartview={FitH},    
    pdftitle={My title},    
    pdfauthor={Author},     
    pdfsubject={Subject},   
    pdfcreator={Creator},   
    pdfproducer={Producer}, 
    pdfkeywords={keyword1} {key2} {key3}, 
    pdfnewwindow=true,      
    colorlinks=true,       
    linkcolor=red,          
    citecolor=cyan,        
    filecolor=magenta,      
    urlcolor=darkgreen,           
    linktocpage=true
}

\allowdisplaybreaks

\DeclareSymbolFontAlphabet{\mathrsfs}{rsfs}
\DeclareMathAlphabet{\mathcal}{OMS}{cmsy}{m}{n}

\newcommand{\beq}{\begin{equation}}
\newcommand{\eeq}{\end{equation}} 

\newcommand{\ud}{\mathrm{d}}
\newcommand{\ui}{\mathrm{i}}

\begin{document}

\title{Gravitational-wave tail effects to quartic non-linear order}

\author{Tanguy Marchand}\email{tanguy.marchand@iap.fr}
\affiliation{$\mathcal{G}\mathbb{R}\varepsilon{\mathbb{C}}\mathcal{O}$,
  Institut d'Astrophysique de Paris,\\ UMR 7095, CNRS, Sorbonne
  Universit{\'e}s \& UPMC Univ Paris 6,\\ 98\textsuperscript{bis}
  boulevard Arago, 75014 Paris, France}
\affiliation{Laboratoire APC -- Astroparticule et Cosmologie, \\
Universit{\'e} Paris Diderot Paris 7, 75013 Paris, France}

\author{Luc Blanchet}\email{luc.blanchet@iap.fr}
\affiliation{$\mathcal{G}\mathbb{R}\varepsilon{\mathbb{C}}\mathcal{O}$,
  Institut d'Astrophysique de Paris,\\ UMR 7095, CNRS, Sorbonne
  Universit{\'e}s \& UPMC Univ Paris 6,\\ 98\textsuperscript{bis}
  boulevard Arago, 75014 Paris, France}

\author{Guillaume Faye}\email{guillaume.faye@iap.fr}
\affiliation{$\mathcal{G}\mathbb{R}\varepsilon{\mathbb{C}}\mathcal{O}$,
  Institut d'Astrophysique de Paris,\\ UMR 7095, CNRS, Sorbonne
  Universit{\'e}s \& UPMC Univ Paris 6,\\ 98\textsuperscript{bis}
  boulevard Arago, 75014 Paris, France}

\date{\today}

\begin{abstract}
  Gravitational-wave tails are due to the backscattering of linear waves onto
  the space-time curvature generated by the total mass of the matter source.
  The dominant tails correspond to quadratic non-linear interactions and arise
  at the one-and-a-half post-Newtonian (1.5PN) order in the gravitational
  waveform. The ``tails-of-tails'', which are cubic non-linear effects
  appearing at the 3PN order in the waveform, are also known. We derive here
  higher non-linear tail effects, namely those associated with quartic
  non-linear interactions or ``tails-of-tails-of-tails'', which are shown to
  arise at the 4.5PN order. As an application, we obtain at that order the
  complete coefficient in the total gravitational-wave energy flux of compact
  binary systems moving on circular orbits. Our result perfectly agrees with
  black-hole perturbation calculations in the limit of extreme mass ratio of
  the two compact objects.
\end{abstract}


\maketitle

\section{Introduction} 
\label{sec:intro}

The LIGO/VIRGO collaboration reported recently the observation of
gravitational waves from the coalescence of black-hole binary
systems~\cite{GW150914,GW151226}. Both analytic works on the two-body problem
in general relativity and extensive numerical relativity calculations play a
very important role when interpreting and deciphering the gravitational-wave
signals~\cite{3mn,CF94,Bliving14,BuonSathya15}.

Our motivation in the present paper, is to find high-order
post-Newtonian (PN) expressions of the gravitational-wave field
generated by the coalescence of compact binary systems (including
black-hole binaries), in the inspiralling phase preceding the final
merger. Such expressions are important for the data analysis of the
ground-based detectors, as well as for the future space-based eLISA
detector. Moreover, they are used for high accuracy comparisons with
the results of numerical relativity.

The current state of the art regarding the gravitational-wave field
can be summarized by listing the various orders that have been
obtained:\footnote{As usual, $n$PN order terms mean terms going up to
  $\sim c^{-2n}$ relatively to the Newtonian quadrupole formula for
  the wave field, and beyond the Newtonian acceleration for the
  equations of motion. Notice the difference of 2.5PN between the two
  nomenclatures, which originates from the fact that the ``Newtonian''
  quadrupole formula corresponds to a 2.5PN radiation reaction effect
  in the equations of motion.} namely, 3.5PN order for the energy
flux~\cite{BDIWW95,B98tail,BFIJ02,BDEI04}, 3PN for the waveform and
polarizations~\cite{BIWW96,ABIQ04,BFIS08}, and 3.5PN for the dominant
gravitational-wave modes~\cite{FMBI12,FBI15}. Regarding the equations
of motion, we have now reached the 4PN
order~\cite{FS4PN,JaraS12,JaraS13,DJS14,BBBFM16a,BBBFM16b}, while the
elucidation of the previous 3PN order had took quite some
time~\cite{JaraS98,JaraS99,DJSpoinc,DJSdim,BFeom,ABF01,BI03CM,BDE04,itoh1,itoh2,FS3PN}.
See~\cite{Bliving14} for a review, and also for the crucial inclusion
of spin effects, both in the equations of motion and radiation field.

Our general aim is to extend the radiation field calculations up to 4.5PN
order, both for the gravitational-wave energy flux (which rules the orbital
phase evolution) and for the polarization waveforms and modes. The first part
of this program consists in dealing with the long computations of the
multipole moments of the compact binary at that order, notably the mass type
quadrupole moment at the 4PN order. We shall leave these computations for
future work.

The second part, addressed in this paper, consists of investigating the
various non-linear interactions between those multipole moments. The most
crucial contributions are due to the so-called gravitational-wave
tails~\cite{Bo59,BoR66,Th80,BD88,BD92}, which are non-linear effects
(quadratic at least) physically due to the backscattering of linear waves onto
the space-time curvature generated by the mass of the source. The tail effects
imply a non-locality in time (i.e., a dependence on the full past history of
the source) so that they may also be qualified as
\textit{hereditary}~\cite{BD92}. The tails arise dominantly at the 1.5PN order
in the waveform, and can be directly tested by the measurement of PN
parameters in LIGO/VIRGO observations~\cite{GW150914,GW151226}.

The tails generated by tails themselves, the so-called ``tails-of-tails'',
come from a cubic interaction and appear dominantly at the 3PN order in the
waveform~\cite{B98tail}. Their contribution can also be potentially tested by
LIGO/VIRGO observations. At the 4.5PN order, new terms called
``tails-of-tails-of-tails'' arise. They come from a \textit{quartic}
interaction between three mass monopoles $M$ (the ADM mass of the source) and
the mass quadrupole moment of the source, say $M_{ij}$. The aim of the present
paper is to compute this quartic multipole interaction $M\times M\times
M\times M_{ij}$ in the asymptotic waveform at large distances.

Our calculation is based on the multipolar-post-Minkowskian (MPM) algorithm
for generating non-linear interactions between multipole moments in the region
outside a general isolated matter source~\cite{BD86,B98quad,B98tail}. Our main
result is the so-called ``\textit{radiative}'' mass-type quadrupole moment
observed at large distances from that source (following the definition
of~\cite{Th80}), given as a functional of the ``\textit{source}'' mass-type
quadrupole moment $M_{ij}$. From that result, we are able to compute the
complete 4.5PN coefficient in the gravitational-wave energy flux for binaries
of (non-spinning) compact objects moving on circular orbits. However, the 4PN
coefficient remains unknown (except in the test mass limit). We relegate its
computation to future work. The reason why we can nonetheless fully compute
the 4.5PN term stems from the fact that half-integral PN approximations in the
flux are only made up of hereditary tail contributions in the case of circular
orbits. We will find that our 4.5PN term agrees, in the test mass limit for
one of the compact bodies, with the expression derived from black-hole
perturbation
theory~\cite{P93a,TNaka94,Sasa94,TSasa94,TTS96,MST96a,MST96b,MT97,Fuj14PN,Fuj22PN}.

The plan of this paper is as follows. Sec.~\ref{sec:recalls} is dedicated to
presenting the necessary material that comes from the MPM approach to
gravitational-wave generation. We investigate in Sec.~\ref{sec:integr} the
integration formulas needed to perform the required non-linear iterations of
the metric. In Sec.~\ref{sec:quartic}, we obtain the leading $1/r$ part of the
quartic-order metric and deduce from it the radiative quadrupole moment.
Finally, in Sec.~\ref{sec:flux}, we derive the complete 4.5PN coefficient in
the total energy flux of compact binaries for circular orbits. The paper ends
with a short conclusion in Sec.~\ref{sec:concl} and two technical Appendices.
 
\section{Review on the multipolar-post-Minkowskian formalism}
\label{sec:recalls}

The gravitational field in the region outside an isolated matter source,
solving the vacuum Einstein field equations in the exterior of that source, is
written in ``gothic'' Minkowskian deviation form\footnote{We have thus
  $h^{\mu\nu}=\sqrt{-g}g^{\mu\nu}-\eta^{\mu\nu}$, where $\eta^{\mu\nu}$ is the
  (inverse) Minkowski metric, while $g$ and $g^{\mu\nu}$ are the determinant
  and the inverse of the covariant metric $g_{\mu\nu}$. Our signature is
  almost plus, i.e., $\eta_{\mu\nu}=\text{diag}(-1,1,1,1)=\eta^{\mu\nu}$. We
  use Cartesian coordinates and solve iteratively the Einstein field equations
  using the harmonic coordinates condition $\partial_\nu h^{\mu\nu}=0$.} and
formally developed as a post-Minkowskian or non-linearity expansion,
\begin{equation}\label{eq:MPM} 
h^{\mu\nu} = G\,h_1^{\mu\nu} + G^2\,h_2^{\mu\nu} + G^3\,h_3^{\mu\nu} +
G^4\,h_4^{\mu\nu} + \mathcal{O}\left(G^5\right)\,,
\end{equation} 
where the powers of the Newton constant $G$ label the successive
approximations. Then, each of the post-Minkowskian coefficients is computed as
a non-linear functional of two infinite sets of time varying
symmetric-trace-free (STF) multipole moments, of mass type, $M_L(t)$, and
current type, $S_L(t)$.\footnote{Here, $L = i_1 \cdots i_\ell$ denotes a
  multi-index composed of $\ell$ spatial indices (ranging from 1 to 3); we
  pose $L-1=i_1 \cdots i_{\ell-1}$, and so on; $\partial_L = \partial_{i_1}
  \cdots \partial_{i_\ell}$ is the product of $\ell$ partial derivatives
  $\partial_i \equiv \partial /
  \partial x^i$; similarly, we shall write $x_L = x_{i_1} \cdots x_{i_\ell}$,
  with $x_i=x^i$ being the spatial position, and $n_L = n_{i_1} \cdots
  n_{i_\ell}$ with $n_i=x_i/r$. Symmetrization over indices is denoted by
  $T_{(ij)}=\frac{1}{2}(T_{ij}+T_{ji})$. The STF projection is indicated with
  a hat, e.g., $\hat{n}_L \equiv \text{STF}[n_L]$, or with angular brackets
  $\langle\rangle$ surrounding the relevant indices, e.g., $x_{\langle
    i}v_{j\rangle}=x_{(i}v_{j)}-\frac{1}{3}\delta_{ij}x_kv_k$. The multipole
  moments $M_L$ and $S_L$ are STF, e.g., $M_L=\hat{M}_L=M_{\langle L\rangle}$.
  Time derivatives of the moments are indicated by superscripts $(n)$. We
  often pose $c=1$ and $G=1$.} However, among these moments, the mass monopole
$M$ is identified with the constant ADM mass of the source, whereas the
constant current dipole $S_i$ coincides with the total angular momentum. The
moments $M_L$ and $S_L$ specifically refer here to the so-called ``canonical''
multipole moments as defined in Ref.~\cite{BFIS08}. The
multipolar-post-Minkowskian (MPM) metric~\eqref{eq:MPM} can in principle be
determined to any order by means of the iterative algorithm described in
Sec.~2 of~\cite{B98quad} (see also Sec.~2.3 in the review~\cite{Bliving14}).
The resulting MPM metric represents the most general solution of the vacuum
Einstein field equations outside the source in harmonic coordinates.

In the present paper, we are interested in non-linear interactions between the
mass $M$ and the mass-type quadrupole moment $M_{ij}$ (having $\ell=2$).
Accordingly, we start the iteration with a linearized metric made only of two
pieces, corresponding to the contributions of $M$ and $M_{ij}$. Then, the
quadratic metric involves three terms corresponding to the various possible
interactions between those moments (including with themselves), and so on.
With an obvious notation,
\begin{subequations}\label{eq:h123}
\begin{align}
h_1^{\mu\nu} &= h_{M}^{\mu\nu} + h_{M_{ij}}^{\mu\nu}\,,\\ h_2^{\mu\nu}
&= h_{M^2}^{\mu\nu} + h_{M\times M_{ij}}^{\mu\nu} + h_{M_{ij}\times
  M_{kl}}^{\mu\nu}\,,\\ h_3^{\mu\nu} &= h_{M^3}^{\mu\nu} +
h_{M^2\times M_{ij}}^{\mu\nu} + h_{M\times M_{ij}\times
  M_{kl}}^{\mu\nu} + h_{M_{ij}\times M_{kl}\times
  M_{mn}}^{\mu\nu}\,.\\ &\vdots\nonumber
\end{align}
\end{subequations}
The iteration is stopped when reaching the non-linear level that is aimed for,
which will here be quartic. In fact, at that level, we are interested only in
the $M^3\times M_{ij}$ quartic interactions, involving three masses and one
quadrupole moment, i.e., we look for the term
\begin{equation}\label{eq:quartic} 
h_4^{\mu\nu} = \cdots + h_{M^3\times M_{ij}}^{\mu\nu} + \cdots\,.
\end{equation} 
Using the MPM algorithm~\cite{BD86,B98quad,B98tail}, this term $h_{M^3\times
  M_{ij}}$ in the gravitational field will be obtained by integrating the
associated source term $\Lambda_{M^3\times M_{ij}}$ entering the vacuum
Einstein field equations in harmonic coordinates. Thus, our task will amount
to solving the ordinary d'Alembertian equation
\begin{equation}\label{eq:boxh} 
\Box h_{M^3\times M_{ij}}^{\mu\nu} = \Lambda_{M^3\times
  M_{ij}}^{\mu\nu}\,,
\end{equation} 
with $\Box=\eta^{\mu\nu}\partial_{\mu\nu}$, while imposing the harmonic gauge
condition $\partial_\nu h_{M^3\times M_{ij}}^{\mu\nu}=0$. The source term
$\Lambda_{M^3\times M_{ij}}$ will be a very complicated expression built from
the previous iterations~\eqref{eq:h123}, see e.g., Sec.~2 of~\cite{Bliving14}
for more details.

Several non-linear interactions are already known, starting with all the
interactions $M\times M\times \cdots$ that involve only the mass monopole,
since they simply reconstitute the Schwarzschild metric in harmonic
coordinates. The quadratic piece $h_{M\times M_{ij}}$, which involves the
quadratic tails, was obtained in~\cite{BD92}. The quadratic interaction
between two quadrupole moments, namely $h_{M_{ij}\times M_{kl}}$, was derived
in~\cite{B98quad}. It contains the well known non-linear memory
effect~\cite{B90,Chr91,WW91,Th92,BD92,B98quad,F09,F11}, as well as
``semi-hereditary'' contributions related to the energy and angular momentum
losses by radiation~\cite{BD92}. The cubic piece $h_{M^2\times M_{ij}}$,
containing the cubic ``tails-of-tails'', was investigated in~\cite{B98tail}.
The cubic interactions $h_{M\times M_{ij}\times M_{kl}}$ and $h_{M_{ij}\times
  M_{kl}\times M_{mn}}$ are not known. Finally the quartic piece $h_{M^3\times
  M_{ij}}$ we are interested in contains the ``tails-of-tails-of-tails''
contribution to the metric.

The mass and quadrupole parts in~\eqref{eq:h123} are given explicitly
by
\begin{subequations}\label{eq:hM}
\begin{align}
  h^{00}_{M} &= - 4 r^{-1} M \,,\\ h^{0i}_{M} &= 0 \,,\\ h^{ij}_{M} &=
  0 \,,
\end{align}
\end{subequations}
and
\begin{subequations}\label{eq:hMij}
\begin{align}
h^{00}_{M_{ij}} &= - 2 n_{ab} r^{-3} \left[ 3 M_{ab}(t-r) + 3 r
  M^{(1)}_{ab}(t-r) + r^2 M^{(2)}_{ab}(t-r)
  \right]\,,\\ h^{0i}_{M_{ij}} &= - 2 n_a r^{-2} \left[
  M^{(1)}_{ai}(t-r) + r M^{(2)}_{ai}(t-r) \right]
\,,\\ h^{ij}_{M_{ij}} &= - 2 r^{-1} M^{(2)}_{ij}(t-r)\,,
\end{align}
\end{subequations}
where the quadrupole moment and its time derivatives depend on the retarded
time $t-r$.

At the quadratic level, we have
\begin{subequations}\label{eq:hM2}
\begin{align}
  h^{00}_{M^2} &= - 7r^{-2} M^2 \,,\\ h^{0i}_{M^2} &= 0
  \,,\\ h^{ij}_{M^2} &= - n_{ij} r^{-2} M^2\,,
\end{align}
\end{subequations}
and
\begin{subequations}\label{eq:hMMij}
\begin{align}
h^{00}_{M\times M_{ij}} &= M n_{ab} r^{-4} \left[ -21 M_{ab} -21 r
  M^{(1)}_{ab} + 7 r^2 M^{(2)}_{ab} + 10 r^3 M^{(3)}_{ab} \right]
\nonumber \\ &+ 8 M n_{ab} \int^{+\infty}_1 \ud y \,Q_2 (y)
M^{(4)}_{ab}(t - ry)\,, \\ 
h^{0i}_{M\times M_{ij}} &= M n_{iab} r^{-3} \left[ -M^{(1)}_{ab} - r
  M^{(2)}_{ab} - \frac{1}{3} r^2 M^{(3)}_{ab} \right] \nonumber \\ &+
M n_a r^{-3} \left[ -5 M^{(1)}_{ai} - 5 r M^{(2)}_{ai} + \frac{19}{3}
  r^2 M^{(3)}_{ai} \right] \nonumber \\ &+ 8 M n_a \int^{+\infty}_1 \ud
y \,Q_1 (y) M^{(4)}_{ai} (t - ry) \,, \\
h^{ij}_{M\times M_{ij}} &= M n_{ijab} r^{-4} \left[ -\frac{15}{2}
  M_{ab} - \frac{15}{2} r M^{(1)}_{ab} - 3 r^2 M^{(2)}_{ab} -
  \frac{1}{2} r^3 M^{(3)}_{ab} \right] \nonumber \\ &+ M \delta_{ij}
n_{ab} r^{-4} \left[ -\frac{1}{2} M_{ab} - \frac{1}{2} r M^{(1)}_{ab}
  - 2 r^2 M^{(2)}_{ab} - \frac{11}{6} r^3 M^{(3)}_{ab} \right]
\nonumber \\ &+ M n_{a(i} r^{-4} \left[ 6 M_{j)a} + 6 r M^{(1)}_{j)a}
  + 6 r^2 M^{(2)}_{j)a} + 4 r^3 M^{(3)}_{j)a} \right] \nonumber \\ &+
M r^{-4} \left[ - M_{ij} - rM^{(1)}_{ij} - 4 r^2 M^{(2)}_{ij} -
  \frac{11}{3} r^3 M^{(3)}_{ij} \right] \nonumber \\ &+ 8 M
\int^{+\infty}_1 \ud y \,Q_0 (y) M^{(4)}_{ij} (t - ry) \,.
\end{align}
\end{subequations}
In the ``instantaneous'' terms, the quadrupole moment is always evaluated at
instant $t-r$. The tail integrals \textit{stricto sensu}, which depend
hereditarily on all past values of the quadrupole moment (evaluated at earlier
time $t-r y$ with $y\geqslant 1$), also contain some specific integration
kernel, which turns out to be the Legendre function of the second kind
$Q_\ell(y)$. The most relevant form of the Legendre function for the present
purpose is displayed in Eq.~\eqref{eq:Qell}. See Ref.~\cite{B98quad} for the
calculation of the lengthy interactions $M_{ij}\times M_{kl}$.

At the cubic level, we have
\begin{subequations}\label{eq:hM3}
\begin{align}
  h^{00}_{M^3} &= - 8 r^{-3} M^3 \,,\\ h^{0i}_{M^3} &= 0
  \,,\\ h^{ij}_{M^3} &= 0\,,
\end{align}
\end{subequations}
while the expressions of the tails-of-tails $M^2\times M_{ij}$ are provided in
Ref.~\cite{B98tail} at the leading asymptotic order in the distance to the
source, when $r\to+\infty$ with $t-r$ fixed. Here however, having in view the
next iteration to compute the tails-of-tails-of-tails, we shall need first to
generalize the latter result to the whole space, i.e., to obtain the
tails-of-tails at any distance $r$ (larger than the size of the source). Thus,
it is worth supplying some more details on the calculation of tails-of-tails.

The source is the sum of a local or instantaneous part, and of an hereditary
or tail part:
\begin{equation}\label{eq:Lambda3} 
\Lambda^{\mu\nu}_{M^2\times M_{ij}} = I^{\mu\nu}_{M^2\times M_{ij}} +
T^{\mu\nu}_{M^2\times M_{ij}}\,.
\end{equation} 
The hereditary part of that cubic source is merely due to the interaction
between $M$ and the tail integrals present in Eqs.~\eqref{eq:hMMij}. We have
\begin{subequations}\label{eq:I3}
\begin{align}
I^{00}_{M^2\times M_{ij}} &= M^2 n_{ab} r^{-7} \biggl[ -516 M_{ab} -
  516 r M^{(1)}_{ab} - 304 r^2 M^{(2)}_{ab} \nonumber \\ &\qquad\qquad
  - 76 r^3 M^{(3)}_{ab} + 108 r^4 M^{(4)}_{ab} + 40 r^5 M^{(5)}_{ab}
  \biggr] \,, \\
I^{0i}_{M^2\times M_{ij}} &= M^2\hat{n}_{iab} r^{-6} \biggl[ 4
  M^{(1)}_{ab} + 4 r M^{(2)}_{ab} - 16 r^2 M^{(3)}_{ab} + \frac{4}{3}
  r^3 M^{(4)}_{ab} - \frac{4}{3} r^4 M^{(5)}_{ab} \biggr] \nonumber
\\ &+ M^2n_a r^{-6} \biggl[ -\frac{372}{5} M^{(1)}_{ai} -
  \frac{372}{5} r M^{(2)}_{ai} -\frac{232}{5} r^2 M^{(3)}_{ai}
  \nonumber \\ &\qquad\qquad- \frac{84}{5} r^3 M^{(4)}_{ai} +
  \frac{124}{5} r^4 M^{(5)}_{ai} \biggr] \,, \\
I^{ij}_{M^2\times M_{ij}} &= M^2\hat{n}_{ijab} r^{-5} \biggl[
  -190 M^{(2)}_{ab} - 118 r M^{(3)}_{ab} 
  - \frac{92}{3} r^2 M^{(4)}_{ab} - 2 r^3
   M^{(5)}_{ab} \biggr] \nonumber \\
  &+ M^2\delta_{ij} n_{ab} r^{-5} \biggl[ \frac{160}{7}
  M^{(2)}_{ab} + \frac{176}{7} r M^{(3)}_{ab}
  - \frac{596}{21}
  r^2 M^{(4)}_{ab} - \frac{160}{21} r^3 M^{(5)}_{ab} \biggr] \nonumber \\
  &+ M^2\hat{n}_{a(i} r^{-5} \biggl[ -\frac{312}{7}
  M^{(2)}_{j)a} - \frac{248}{7} r M^{(3)}_{j)a} 
  + \frac{400}{7}
  r^2 M^{(4)}_{j)a} + \frac{104}{7} r^3 M^{(5)}_{j)a} \biggr] \nonumber \\
  &+ M^2r^{-5} \biggl[ -12 M^{(2)}_{ij} - \frac{196}{15}
  r M^{(3)}_{ij} - \frac{56}{5} r^2
  M^{(4)}_{ij} - \frac{48}{5} r^3 M^{(5)}_{ij}
  \biggr]\,,
\end{align}\end{subequations}
and
\begin{subequations}\label{eq:T3}
\begin{align}
T^{00}_{M^2\times M_{ij}} &= M^2 n_{ab} r^{-3} \int^{+\infty}_1 \ud y
\biggl[ 96 Q_0 M^{(4)}_{ab} + \left( \frac{272}{5} Q_1 + \frac{168}{5}
  Q_3 \right) r M^{(5)}_{ab} + 32 Q_2 r^2 M^{(6)}_{ab} \biggr]\,, \\
T^{0i}_{M^2\times M_{ij}} &= M^2\hat{n}_{iab} r^{-3} \int^{+\infty}_1
\ud y \biggl[ - 32 Q_1 M^{(4)}_{ab} + \left( -\frac{32}{3} Q_0 +
  \frac{8}{3} Q_2 \right) r M^{(5)}_{ab} \biggr] \nonumber \\ &+
M^2n_a r^{-3} \int^{+\infty}_1 \ud y \biggl[ \frac{96}{5} Q_1
  M^{(4)}_{ai} + \left( \frac{192}{5} Q_0 + \frac{112}{5} Q_2 \right)
  r M^{(5)}_{ai} + 32 Q_1 r^2 M^{(6)}_{ai} \biggr]\,, \\
T^{ij}_{M^2\times M_{ij}} &= M^2\hat{n}_{ijab} r^{-3} \int^{+\infty}_1
\ud y \biggl[ - 32 Q_2 M^{(4)}_{ab} + \left(- \frac{32}{5} Q_1 -
  \frac{48}{5} Q_3 \right) r M^{(5)}_{ab} \biggr] \nonumber \\ &+
M^2\delta_{ij} n_{ab} r^{-3} \int^{+\infty}_1 \ud y \biggl[ -
  \frac{32}{7} Q_2 M^{(4)}_{ab} + \left( - \frac{208}{7} Q_1 +
  \frac{24}{7} Q_3 \right) r M^{(5)}_{ab} \biggr] \nonumber \\ &+
M^2\hat{n}_{a(i} r^{-3} \int^{+\infty}_1 \ud y \biggl[ \frac{96}{7}
  Q_2 M^{(4)}_{j)a} + \left( \frac{2112}{35} Q_1 - \frac{192}{35} Q_3
  \right) r M^{(5)}_{j)a} \biggr] \nonumber \\ &+ M^2r^{-3}
\int^{+\infty}_1 \ud y \biggl[ \frac{32}{5} Q_2 M^{(4)}_{ij} + \left(
  \frac{1536}{75} Q_1 - \frac{96}{75} Q_3 \right) r M^{(5)}_{ij} + 32
  Q_0 r^2 M^{(6)}_{ij} \biggr]\,.
\end{align}\end{subequations}
As in the tail integrals of Eqs.~\eqref{eq:hMMij}, in the above equations, the
Legendre functions are evaluated at $y$ and the quadrupole moments are
evaluated at $t-ry$.

We now apply the MPM algorithm~\cite{BD86,B98quad,B98tail} to compute the
$M^2\times M_{ij}$ metric. We first define a particular solution of the
d'Alembertian equation $\Box h_{M^2\times M_{ij}} = \Lambda_{M^2\times
  M_{ij}}$ for $\Lambda_{M^2\times M_{ij}}$ given by
Eqs.~\eqref{eq:Lambda3}--\eqref{eq:T3}, as
\begin{equation}
u^{\mu\nu}_{M^2\times M_{ij}} = \mathop{\mathrm{FP}}_{B=0}
\,\Box^{-1}_R \biggl[ \left( \frac{r}{r_0}\right)^B
  \Lambda^{\mu\nu}_{M^2\times M_{ij}} \biggr]\,.\label{eq:umunu}
\end{equation}
Here, $\Box^{-1}_R$ denotes the usual three-dimensional retarded integral. The
source term is regularized by means of a multiplying factor $(r/r_0)^B$, with
$B$ being a complex number and $r_0$ denoting a certain constant length scale.
The object~\eqref{eq:umunu} is defined by analytic continuation in
$B\in\mathbb{C}$ (over the complex plane deprived of some isolated points).
The finite part operation at $B=0$ (in short FP$_{B=0}$) selects the zero-th
order coefficient of the Laurent expansion when $B\to 0$. As a result of this
definition, we have $\Box u_{M^2\times M_{ij}}=\Lambda_{M^2\times M_{ij}}$ (so
$u_{M^2\times M_{ij}}$ is indeed a particular solution). Moreover,
$u_{M^2\times M_{ij}}$ has a multipolar structure similar to that of the
source term.

In the next stage, we compute the divergence of~\eqref{eq:umunu}, namely
$w^{\mu}_{M^2\times M_{ij}}=\partial_\nu u^{\mu\nu}_{M^2\times M_{ij}}$. Using
the fact that the divergence of the non-linear source term is zero by virtue
of the Bianchi identities, i.e., $\partial_\nu \Lambda^{\mu\nu}_{M^2\times
  M_{ij}}=0$, which can be checked by a direct calculation on the
expressions~\eqref{eq:I3}--\eqref{eq:T3}, we get
\begin{equation}
w^{\mu}_{M^2\times M_{ij}} = \mathop{\mathrm{FP}}_{B=0} \,\Box^{-1}_R
\biggl[ B \left( \frac{r}{r_0}\right)^B \frac{n^i}{r}\,\Lambda^{\mu
    i}_{M^2\times M_{ij}} \biggr]\,,\label{eq:wmu}
\end{equation}
where the factor $B$ comes from the differentiation of the regularization
factor $(r/r_0)^B$. Therefore, the solution~\eqref{eq:umunu} is not
divergence-free in general but one can prove that, because of the factor $B$,
its divergence is a homogeneous retarded solution of the d'Alembertian
equation, i.e., $\Box w^{\mu}_{M^2\times M_{ij}}=0$. It is then
straightforward to find a correction term $v^{\mu\nu}_{M^2\times M_{ij}}$
satisfying at once $\Box v^{\mu\nu}_{M^2\times M_{ij}}=0$ and $\partial_\nu
v^{\mu\nu}_{M^2\times M_{ij}}=-w^{\mu}_{M^2\times M_{ij}}$. The
equations~(2.11)--(2.12) of~\cite{B98quad} allow one to construct
algorithmically $v^{\mu\nu}_{M^2\times M_{ij}}$ starting from
$w^{\mu}_{M^2\times M_{ij}}$. Finally, the MPM solution of the Einstein field
equations in harmonic coordinates ($\partial_\nu h^{\mu\nu}_{M^2\times
  M_{ij}}=0$) reads
\begin{equation}
h^{\mu\nu}_{M^2\times M_{ij}} = u^{\mu\nu}_{M^2\times M_{ij}} +
v^{\mu\nu}_{M^2\times M_{ij}}\,.\label{eq:hmunu}
\end{equation}

Following this algorithm, Ref.~\cite{B98tail} obtained the dominant terms of
the solution $h_{M^2\times M_{ij}}$ at infinity, when $r\to+\infty$ with
$t-r=$ const. When looking only for the dominant asymptotic behaviour of the
solution, we dispose of a simplified version of the algorithm given in the
Appendix~B of~\cite{B98tail}. The result is
\begin{subequations}\label{eq:asympt}
\begin{align}
h^{00}_{M^2\times M_{ij}} &= \frac{M^2 n_{ab}}{r} \int^{+\infty}_0
\ud\tau\, M^{(5)}_{ab} \biggl[ -4 \ln^2 \left( \frac{\tau}{2r} \right) -4
\ln \left( \frac{\tau}{2r} \right) \nonumber\\&\qquad\qquad +
\frac{116}{21} \ln \left( \frac{\tau}{2r_0} \right) -
\frac{7136}{2205} \biggr] +
\mathcal{O}\left(\frac{1}{r^{2-\epsilon}}\right)\,,\\
h^{0i}_{M^2\times M_{ij}} &= \frac{M^2 \hat{n}_{iab}}{r}
\int^{+\infty}_0 \ud\tau\, M^{(5)}_{ab} \biggl[ -\frac{2}{3} \ln \left(
\frac{\tau}{2r} \right) -\frac{4}{105} \ln \left( \frac{\tau}{2r_0}
\right) - \frac{716}{1225} \biggr] \nonumber\\&+ \frac{M^2 n_a}{r}
\int^{+\infty}_0 \ud\tau\, M^{(5)}_{ai} \biggl[ -4 \ln^2 \left(
\frac{\tau}{2r} \right) - \frac{18}{5} \ln \left( \frac{\tau}{2r}
\right) \nonumber\\&\qquad\qquad + \frac{416}{75} \ln \left(
\frac{\tau}{2r_0} \right) - \frac{22724}{7875} \biggr] +
\mathcal{O}\left(\frac{1}{r^{2-\epsilon}}\right)\,,\\
h^{ij}_{M^2\times M_{ij}} &= \frac{M^2 \hat{n}_{ijab}}{r}
\int^{+\infty}_0 \ud\tau\, M^{(5)}_{ab} \biggl[ - \ln \left(
\frac{\tau}{2r} \right) - \frac{191}{210} \biggr] \nonumber\\&+
\frac{M^2 \delta_{ij} n_{ab}}{r} \int^{+\infty}_0 \ud\tau\, M^{(5)}_{ab}
\biggl[ -\frac{80}{21} \ln \left( \frac{\tau}{2r} \right) -
\frac{32}{21} \ln \left( \frac{\tau}{2r_0} \right) - \frac{296}{35}
\biggr] \nonumber\\&+ \frac{M^2 \hat{n}_{a(i}}{r} \int^{+\infty}_0
\ud\tau\, M^{(5)}_{j)a} \biggl[ \frac{52}{7} \ln \left( \frac{\tau}{2r}
\right) + \frac{104}{35} \ln \left( \frac{\tau}{2r_0} \right) +
\frac{8812}{525} \biggr] \nonumber\\&+ \frac{M^2}{r} \int^{+\infty}_0
\ud\tau\, M^{(5)}_{ij} \biggl[ -4 \ln^2 \left( \frac{\tau}{2r} \right) -
\frac{24}{5} \ln \left( \frac{\tau}{2r} \right)
\nonumber\\&\qquad\qquad + \frac{76}{15} \ln \left( \frac{\tau}{2r_0}
\right) - \frac{198}{35} \biggr] +
\mathcal{O}\left(\frac{1}{r^{2-\epsilon}}\right)\,.
\end{align}\end{subequations}
The quadrupole moments in the integrands are evaluated at instant $t-r-\tau$.
The notation $\mathcal{O}(r^{\epsilon-2})$ [which could even be
$o(r^{\epsilon-2})$], with $0<\epsilon\ll 1$, is simply to account for the
presence of logarithms of $r$ in the expansion at infinity, as the remainder
is really made of a sum of some $\mathcal{O}(r^{-2}\ln^a r)$, with $a=0,1$
in~\eqref{eq:asympt}.

In this paper, we shall generalize Eqs.~\eqref{eq:asympt} to the whole space
(the complete expressions will be too long to be displayed) and use them,
along with many other interaction terms involving the lower order
metrics~\eqref{eq:hM}--\eqref{eq:hM3}, to construct the quartic source term
$\Lambda_{M^3\times M_{ij}}$, which is precisely the source of the looked-for
tails-of-tails-of-tails contribution~\eqref{eq:quartic}. An important check of
this calculation will be to make sure that
$\partial_\nu\Lambda^{\mu\nu}_{M^3\times M_{ij}}=0$. At that stage, following
again the MPM algorithm, we shall define
\begin{equation}\label{eq:uquartic}
u^{\mu\nu}_{M^3\times M_{ij}} = \mathop{\mathrm{FP}}_{B=0}
\,\Box^{-1}_R \biggl[ \left( \frac{r}{r_0}\right)^B
  \Lambda^{\mu\nu}_{M^3\times M_{ij}} \biggr]\,.
\end{equation}
Since we are at the final iteration step, we shall be content with the leading
asymptotic behaviour at infinity, when $r\to+\infty$ with $t-r=$ const, of the
solution being constructed. There remains to compute the divergence
$w_{M^3\times M_{ij}}$ of \eqref{eq:uquartic}, to check that it is a
homogeneous retarded solution of the d'Alembertian equation, at leading order
in $1/r$, and finally to add the correcting piece $v_{M^3\times M_{ij}}$
ensuring that the harmonic-coordinate condition is satisfied. In the large $r$
limit, Eqs.~(B.4)--(B.5) of~\cite{B98tail} are the relevant formulas to go
from $w_{M^3\times M_{ij}}$ to $v_{M^3\times M_{ij}}$. In the end, our quartic
metric will be built as
\begin{equation}
h^{\mu\nu}_{M^3\times M_{ij}} = u^{\mu\nu}_{M^3\times M_{ij}} +
v^{\mu\nu}_{M^3\times M_{ij}}\,.\label{eq:hmunuquartic}
\end{equation}
The leading $1/r$ behaviour of $h_{M^3\times M_{ij}}$ (actually containing
also $\ln^2r/r$ and $\ln r/r$ terms) will be shown in Eqs.~\eqref{eq:hM3Mij}
below. The physical radiative quadrupole moment at infinity will be extracted
from that metric in Eq.~\eqref{eq:radquad}. As we shall see, the quartic
tails-of-tails-of-tails represent dominantly a 4.5PN effect in the waveform,
which will be specialized to compute the total energy flux generated by
compact binary sources moving on circular orbits in Eq.~\eqref{eq:Foddx}.

\section{Formulas to compute quartic non-linearities} 
\label{sec:integr}

In this section, we present the basic integration formulas (extending notably
the Appendix A of~\cite{B98tail}) allowing for the integration of the cubic
and quartic non-linearities in essentially analytic closed form. We are
looking for the retarded solution of a certain d'Alembertian equation whose
source term, which represents a generic term in Eqs.~\eqref{eq:T3}, is
hereditary:
\begin{equation}
\Box \mathop{\Psi}_{k,m}\!\!{}_L = \hat{n}_L \,r^{-k}
\int_{1}^{+\infty} \ud y \,V_m(y) F(t - ry)\,,\label{eq:dalembertian}
\end{equation}
where $\hat{n}_L$ is a STF product of $\ell$ unit vectors (with $L=i_1\cdots
i_\ell$), $F(u)$ a smooth function of the retarded time that is identically
zero in the remote past, i.e., $F\in\mathcal{C}^\infty(\mathbb{R})$ and
$F(u)=0$ for $u\leqslant -\mathcal{T}$ (with $-\mathcal{T}$ being a fixed
instant in the past), and where $V_m(y)$ is a generic function belonging to
the following $m$-dependent class:
\begin{align}
\mathscr{V}_m = \Bigl\{&V(y) \in \mathcal{C}^{\infty}( ] 1, +\infty[)
    \mid \exists\ a \geqslant 0,\ b \geqslant 0\ \text{such
      that}\nonumber\\ &V(y) \mathop{=}_{y \to +\infty}
    \mathcal{O}\left[y^{-(m+1)}\ln^a(y)\right] \ \text{and} \ V(y)
    \mathop{=}_{y \to 1^+} \mathcal{O}\left[\ln^b(y-1)\right]
    \Bigr\}\,.\label{eq:defFm}
\end{align}
We see that the integer $m$ basically specifies the behaviour of our
$\mathscr{V}_m$-type functions when $y\to+\infty$, while those functions are
assumed to be integrable when $y\to 1^+$. A typical function belonging to the
class $\mathscr{V}_m$ is the Legendre function of the second kind $Q_m(y)$,
given by~\eqref{eq:Qell} below. Then, for $V_m \in \mathscr{V}_m$, we define
the retarded multipolar solution of~\eqref{eq:dalembertian} as
\begin{equation}
\mathop{\Psi}_{k,m}\!\!{}_L = \mathop{\mathrm{FP}}_{B=0} \,\Box^{-1}_R
\biggl[ \hat{n}_L \left( \frac{r}{r_0}\right)^B r^{-k}
  \int_{1}^{+\infty} \ud y \,V_m(y) F(t - ry)
  \biggr]\,.\label{eq:solgen}
\end{equation}
Following the prescriptions~\eqref{eq:umunu} or~\eqref{eq:uquartic}, this
solution is defined by analytic continuation in $B\in\mathbb{C}$ as the finite
part (FP) in the Laurent expansion when $B\to 0$ of the usual inverse retarded
integral $\Box^{-1}_R$ acting on the source regularized by means of the
inserted factor $(r/r_0)^B$. The arbitrary constant scale $r_0$ will cancel
out from our final physical result in Sec.~\ref{sec:flux}.

\subsection{Explicit closed-form representations of the solution} 
\label{sec:explicit}

We shall now present explicit forms for the general
solution~\eqref{eq:solgen}, i.e., analytic closed-form representations for
${}_{k,m}\Psi_L$. Such representations are indispensable when implementing in
practice the non-linear iterative construction of the metric. In order to get
the full cubic metric, whose source is the sum of~\eqref{eq:I3}
and~\eqref{eq:T3}, we need to distinguish several cases.\footnote{Here we do
  not discuss the integration of the instantaneous terms~\eqref{eq:I3} which
  is comparatively much simpler than that of the hereditary terms and can be
  dealt with the formulas in Appendix~A of~\cite{B98quad}.}

\subsubsection{Case where $k=1$, $\ell\geqslant 0$ and $m \geqslant 0$}
\label{sec:case1}

This case has been already investigated in Ref.~\cite{B98tail} when $V_m =
Q_m$ is the Legendre function of the second kind. The result extends naturally
to all $V_m \in \mathscr{V}_m$:
\begin{equation}
\mathop{\Psi}_{1,m}\!\!{}_L = \hat{n}_L \int_{1}^{+\infty} \ud s
F^{(-1)}(t-rs) \left[ Q_\ell (s) \int_{1}^s \ud y \,V_m(y) \,\frac{\ud
    P_\ell}{\ud y}(y) + P_\ell(s) \int_{s}^{+\infty} \ud y \,V_m(y)
  \,\frac{\ud Q_\ell}{\ud y} (y)\right]\,. \label{eq:kequal1}
\end{equation}
Since $F(u)$ is identically zero in the past (before some given finite instant
$-\mathcal{T}$), we define $F^{(-1)}(u)$ to be the anti-derivative of $F$ that
is also identically zero in the past. Here, $P_\ell(y)$ is the usual Legendre
polynomial. The Legendre function $Q_{\ell} (y)$ of the second kind, with a
branch cut from $y=-\infty$ to $y=1$ in the complex plane, takes the explicit
form~\cite{GR}\footnote{Two other forms useful in the present context are
$$Q_{\ell} (y) = \frac{1}{2} \int_{-1}^1 \ud x\,\frac{P_\ell (x)}{y-x}
  = \frac{1}{2^{\ell+1}} \int_{-1}^1 \ud
  z\,\frac{(1-z^2)^\ell}{(y-z)^{\ell+1}}\,.$$
}
\begin{equation}\label{eq:Qell}
Q_{\ell} (y) = \frac{1}{2} P_\ell (y) \ln \left(\frac{y+1}{y-1}
\right)- \sum^{\ell}_{j=1} \frac{1}{j} \,P_{\ell -j}(y) P_{j-1}(y)\,.
\end{equation}
We recall that this function behaves as $Q_\ell(y)\sim y^{-\ell-1}$ when
$y\to+\infty$, and that its leading expansion when $y\to 1^+$ reads
\begin{equation}
Q_\ell(y)=-\frac{1}{2}\ln\left(\frac{y-1}{2}\right)-H_\ell +
\mathcal{O}(y-1)\,, \label{eq:Qelly1}
\end{equation}
where $H_{\ell}=\sum_{j=1}^{\ell} \frac{1}{j}$ denotes the usual harmonic
number.

\subsubsection{Case where $k=2$, $\ell\geqslant 0$ and $m \geqslant 0$}

For $k=2$, and still $V_m \in \mathscr{V}_m$ with $m \geqslant 0$, we start
with the formula~(D5) of Appendix D in Ref.~\cite{BD86}, which yields, for the
case at hands:
\begin{align}
\mathop{\Psi}_{2,m}\!\!{}_L &= - \frac{\hat{n}_L}{2r}
\int_{-\infty}^{t-r}\ud \xi
\int_{\frac{t-r-\xi}{2}}^{\frac{t+r+\xi}{2}} \frac{\ud w}{w}
\int_1^{+\infty} \ud x \,V_m(x)\,F\big[\xi - (x-1)w\big] \nonumber
\\ &\qquad\qquad\qquad \times P_\ell\left[1 - \frac{(t-r-\xi)(t+r-\xi
    - 2w)}{2rw}\right]\,.
\end{align}
Now, we define new variables $(\xi,w)\rightarrow (y,z)$ by $\xi - (x-1)w = t -
r y$ and $z = 1 - \frac{(t-r-\xi)(t+r-\xi - 2w)}{2rw}$. With these variables
we get
\begin{align}
\mathop{\Psi}_{2,m}\!\!{}_L &= - \frac{\hat{n}_L}{2}
\int_{1}^{+\infty}\ud x \,V_m(x) \int_{1}^{+\infty} \ud y \,F(t-ry)
\int_{-1}^1 \ud z \,\frac{P_\ell(z)}{\sqrt{(xy-z)^2 -
    (x^2-1)(y^2-1)}}\,.
\end{align}
By virtue of the mathematical formula (A.5) of~\cite{B98tail}\footnote{Namely,
$$\frac{1}{2} \int^1_{-1} \frac{\ud z \,P_\ell (z)}{\sqrt{(xy - z)^2 -
      (x^2 - 1) (y^2 - 1)}} = \left\{\begin{array}{l}P_\ell (x)
  \,Q_\ell (y) ~~\text{when}~~ 1 < x \leqslant y \,, \\[0.2cm] P_\ell
  (y) \,Q_\ell (x) ~~\text{when}~~ 1 < y \leqslant x\,. \end{array}
  \right. $$} we obtain:
\begin{equation}
\label{eq:kequal2}
\mathop{\Psi}_{2,m}\!\!{}_L = - \hat{n}_L \int_{1}^{+\infty} \ud s
\,F(t-rs) \left[ Q_\ell (s) \int_{1}^s\ud y \,V_m(y) P_\ell(y) +
  P_\ell(s) \int_{s}^{+\infty} \ud y \,V_m(y) Q_\ell(y)\right]\,,
\end{equation}
which has a structure similar to that of the solution~\eqref{eq:kequal1}.

\subsubsection{Case where $k\geqslant 2$, $\ell \geqslant k-2$ and $m \geqslant k-2$}

To deal with this case, it is convenient to introduce, given some positive
integer $p$ and some function $V_m\in \mathscr{V}_m$, the $p$-th anti-derivative
$V_m^{(-p)}(y)$ of $V_m$ that vanishes when $y=1$, together with all its
derivatives of orders smaller than $p$. Namely, we define
\begin{equation}\label{eq:primVp}
V_m^{(-p)}(y) = \int_{1}^y \ud x \,V_m(x) \,\frac{(y-x)^{p-1}}{(p-1)!}\,,
\end{equation}
and adopt the convention that $V_m^{(0)}(y) = V_m(y)$. Such a choice is indeed
meaningful for functions $V_m$ that satisfy the characteristic properties of
the class $\mathscr{V}_m$. Now, for any $\ell \geqslant k-2$ and $m \geqslant
k-2$, we have shown that the solution ${}_{k,m}\Psi_L$ is given by
\begin{align}
\mathop{\Psi}_{k,m}\!\!{}_L &= - \hat{n}_L \int_{1}^{+\infty} \ud s
\,F^{(k-2)}(t-rs) \biggl[ Q_\ell (s) \int_{1}^s\ud y \,V_m^{(-k+2)}(y)
  P_\ell(y) \nonumber \\ & \qquad\qquad\qquad + P_\ell(s)
  \int_{s}^{+\infty} \ud y \,V_m^{(-k+2)}(y)
  Q_\ell(y)\biggr]\,, \label{eq:kgeq2}
\end{align}
which appears to be an interesting generalization of Eq.~\eqref{eq:kequal2}
corresponding to the case $k=2$. Notice, however, that the latter
formula~\eqref{eq:kgeq2} is not valid in the case $k=1$. This case has to be treated
separately using the result~\eqref{eq:kequal1}. The proof of
Eq.~\eqref{eq:kgeq2} goes by induction on the integer $k \geqslant 2$, and is
relegated to Appendix~\ref{app:proof1}.

\textit{Stricto sensu}, we are not allowed to use Eq.~\eqref{eq:kgeq2} when
$m=0$, $k=3$, $\ell=2$, which corresponds to one of the hereditary terms of
the cubic source~\eqref{eq:T3}. However, it happens to be valid also in this
case. Indeed, the proof leading to Eq.~\eqref{eq:kequal2} still holds for
$m=-1$ as all integrals are convergent. Then, to derive the formula for $m=0$,
$k=3$, $\ell=2$, we proceed similarly to the recursion presented in
Appendix~\ref{app:proof1}, by performing an integration by parts and choosing
for $V_0^{(-1)}(y)$ the anti-derivative that vanishes for $y=1$ [see
Eq.~\eqref{eq:intpart}].

\subsubsection{Case $\ell=0$, $k\geqslant 3$ and $m\geqslant k-2$}
\label{sec:case4}

As it turned out, one (and only one) term of the cubic source given by
Eq.~\eqref{eq:T3} does not fall into the previous cases. For this term,
corresponding to the values $\ell=0$, $k=3$ and $m=2$, we need to find another
formula. Fortunately, this can be done by noticing that, when $k=2$,
Eq.~\eqref{eq:kequal2} is true for $\ell=0$, and when $k=3$, most of the
reasonings of the proof of Eq.~\eqref{eq:kgeq2} remain valid. More details are
given in Appendix~\ref{app:proof2}. In the end, for $k=3$ we find
\begin{align}
\mathop{\Psi}_{3,m}\!\!{}_{L=0} =& -
\biggl[\ln\left(\frac{r}{r_0}\right)+1\biggr]\frac{F(t-r)}{r}
\int_{1}^{+\infty} \ud y \,V_m(y) \nonumber \\ & + \int_{1}^{+\infty}
\ud s \,F^{(1)}(t-rs) \left(Q_0(s) \int_{1}^s\ud y \,(y+1) V_m(y)
\right. \nonumber \\ &\left.\qquad\qquad+ \int_{s}^{+\infty}\ud y
\left[(y+1)Q_0(y) + \ln \left(\frac{y-1}{s-1} \right) \right] V_m(y)
\right) \,,\label{eq:k3}
\end{align}
where we recall that $Q_0(y)=\frac{1}{2}\ln(\frac{y+1}{y-1})$. Observe the
first appearance of the logarithm of $r$, in the first term of
Eq.~\eqref{eq:k3}, due to the presence of a pole in the original integral when
$B\to 0$. As a result the formula~\eqref{eq:k3} explicitly depends on the
arbitrary scale $r_0$. It will be interesting to study later the fate of such
scale which must disappear from physical results. With Eq.~\eqref{eq:k3}, we
have in hands enough material to integrate explicitly all the cubic hereditary
source terms given by~\eqref{eq:T3} --- the integration of the instantaneous
source terms~\eqref{eq:I3} being the same as for lower orders.

We have also derived a more general formula, valid for $\ell=0$, $k\geqslant
3$ and $m \geqslant k-2$:
\begin{align} 
\mathop{\Psi}_{k,m}\!\!{}_{L=0} =& \frac{(-)^k}{(k-2)!}\Biggl\{
\biggl[\ln\left(\frac{r}{r_0}\right)+H_{k-2}\biggr]\frac{F^{(k-3)}(t-r)}{r}
\int_{1}^{+\infty} \ud y \,V_m(y) \varphi_{k-2}(y) \nonumber \\ &\quad
- \int_{1}^{+\infty} \ud s \,F^{(k-2)}(t-rs) \biggl( Q_0(s)\int_1^s
\ud y\,V_m(y)\,(y+1)^{k-2} \nonumber \\ & \qquad\quad +
\int_{s}^{+\infty}\ud y\,V_m(y) \biggl[(y+1)^{k-2} Q_0(y) +
  \varphi_{k-2}(y)\ln\left(\frac{y-1}{s-1}\right)\biggr]\biggr)\nonumber
\\ &\quad - \sum_{i=1}^{k-3} (-)^{k+i}\frac{(k-3-i)!}{r^{k-1-i}}
\int_{1}^{+\infty} \ud y \,V_m(y) \,\varphi_i(y) \,F^{(i-1)}(t-ry)
\Biggr\}\,,\label{eq:plusgen}
 \end{align}
 where we have posed $\varphi_i(y)=\frac{1}{2}[(y+1)^i-(y-1)^i]$, to ease the
 notation, and $H_{k-2}=\sum_{j=1}^{k-2} \frac{1}{j}$. Notice the last term
 in~\eqref{eq:plusgen}, which is absent from Eq.~\eqref{eq:k3} and constitutes
 an additional contribution here for $k\geqslant 4$. The proofs
 of~\eqref{eq:k3} and~\eqref{eq:plusgen} are presented in
 Appendix~\ref{app:proof2}.

 For all the previous formulas in this section, we have verified explicitly
 that the original d'Alembertian equation~\eqref{eq:dalembertian} is satisfied
 and that the leading asymptotic behaviour, for $r \to \infty$ with $t-r$
 constant, is in complete agreement with Eqs.~(A.13) and (A.19) of
 Ref.~\cite{B98tail} in the particular case where $V_m(y) = Q_m(y)$.

\subsection{Asymptotic expansion at future null infinity}
\label{sec:asympt}

We now present other formulas, going beyond those investigated in
Ref.~\cite{B98tail}, for studying the leading order in the asymptotic
expansion when $r \to +\infty$ with $t-r$ constant, and which will enable us
to control the asymptotic behaviour of the metric in the last stage of our
iteration, at the quartic level.

\subsubsection{Case $k = 1$, $m \geqslant 0$ and $\ell\geqslant 0$}

From the result~\eqref{eq:kequal1} it is straightforward to see (\textit{cf}
Eq.~(A.7) of~\cite{B98tail}) that, to leading order at future null infinity
($r\to +\infty$ with $t-r=$ const):
\begin{equation}
\mathop{\Psi}_{1,m}\!\!{}_{L} = \frac{\hat{n}_L}{r} \int_{0}^{+\infty}
\ud\tau\, F^{(-1)} (t- r- \tau) \int_{1 + \tau/r}^{+\infty} \ud x
\,V_m(x)\frac{\ud Q_\ell}{\ud x}(x) +
\mathcal{O}\left(\frac{1}{r^{2-\epsilon}}\right)\,. \label{eq:behaviourkequal1}
\end{equation}
We remind that the neglected terms in $\mathcal{O}(r^{\epsilon-2})$ also
include possible powers of the logarithm of $r$.

\subsubsection{Case $k\geqslant 2$, $\ell \geqslant k-2$, $m \geqslant 0$}

The formulas~(A.10)--(A.17) of~\cite{B98tail} can be extended to any function
$V_m \in \mathscr{V}_m$ by means of the same procedure that was used to get
them in~\cite{B98tail}. We find in that case that
\begin{equation}
\mathop{\Psi}_{k,m}\!\!{}_{L} = -
\mathop{\alpha}_{k,m}\!\!{}_{\ell}\,\frac{\hat{n}_L}{r}
\,F^{(k-3)}(t-r) + \mathcal{O}\left(\frac{1}{r^2}\right)\,.
\end{equation}
The coefficients are given by the following explicit although involved
expressions:
\begin{subequations}\label{eq:alphaC}
\begin{align}
\mathop{\alpha}_{k,m}\!\!{}_{\ell} &= \sum_{i=0}^{k-2}
C_{\ell,i}^{k-2} \int_1^{+\infty} \ud y \,V_m(y)
Q_{\ell-k+2+2i}(y)\,, \label{eq:alphaCa} \\ \text{where}\quad C_{\ell,i}^{k-2} &=
(-)^i\genfrac{(}{)}{0pt}{}{k-2}{i} \frac{(2\ell - 2k + 3+
  2i)!!}{(2\ell + 1 + 2i)!!}  (2\ell - 2k + 5 + 4i)
\,,\label{eq:alphaCb}
\end{align}\end{subequations}
with ${\genfrac{(}{)}{0pt}{}{k-2}{i}}$ denoting the usual binomial
coefficient. One can check that the remaining integral is convergent for any
$V_m\in\mathscr{V}_m$ as long as $\ell \geqslant k-2$, since $Q_\ell(y)\sim
y^{-\ell-1}$ when $y\to+\infty$. Interestingly, the expression~\eqref{eq:alphaCa} for the
${}_{k,m}\alpha_{\ell}$'s may be recast into the more compact form
\begin{equation}
\mathop{\alpha}_{k,m}\!\!{}_{\ell} = (-)^k
\int_1^{+\infty} \ud y \,V_m(y)Q^{(-k+2)}_\ell(y)\,,
\end{equation}
where $Q^{(-k+2)}_\ell(y)$ is the $(k-2)$-th
anti-derivative of $Q_\ell(y)$ that vanishes at $y=+\infty$ with all its
derivatives, i.e.,
\begin{equation}\label{eq:antiderQ}
Q^{(-k+2)}_\ell(y) = - \int_{y}^{+\infty}\ud
z\,Q_\ell(z)\,\frac{(y-z)^{k-3}}{(k-3)!} \,,
\end{equation}
for $k\geqslant 3$, and $Q^{(0)}_\ell(y)\equiv Q_\ell(y)$.

\subsubsection{Case $k \geqslant \ell+3$ and $m \geqslant k-\ell-2$}

Adapting the equations (A.19)--(A.21) from~\cite{B98tail} we readily get
\begin{equation}
\mathop{\Psi}_{k,m}\!\!{}_{L} = - \frac{\hat{n}_L}{r}
\int_{0}^{+\infty} \ud \tau \,F^{(k-2)}(t-r-\tau)
\left[\mathop{\beta}_{k,m}\!\!{}_{\ell} \ln \left(\frac{\tau}{2 r_0}
  \right) + \mathop{\gamma}_{k,m}\!\!{}_{\ell}\right] +
\mathcal{O}\left(\frac{1}{r^2}\right)\,,
\end{equation}
with the explicit coefficients
\begin{subequations}\label{eq:coeffs}
\begin{align}
\mathop{\beta}_{k,m}\!\!{}_{\ell} &= \frac{1}{2} \int_1^{+\infty} \ud
x \,V_m(x) \int_{-1}^1 \ud z \,\frac{(z-x)^{k-3}}{(k-3)!}
P_\ell(z)\,,\\ \mathop{\gamma}_{k,m}\!\!{}_{\ell} &= \frac{1}{2}
\int_1^{+\infty} \ud x \,V_m(x) \int_{-1}^1 \ud z
\,\frac{(z-x)^{k-3}}{(k-3)!}  P_\ell(z) \left[-\ln
  \left(\frac{x-z}{2}\right) + H_{k-3}\right]\,.
\end{align}
\end{subequations}

\subsubsection{Case $k=4$, $\ell=0$ and $m=0$}

So far, we have just extended in a natural way the integration formulas
of~\cite{B98tail} (see also the Appendix A of~\cite{FBI15} for other
formulas). However, in our computation, one extra case must still be dealt
with, corresponding to the values $k=4$, $\ell=0$ and $m=0$. Because $m$
vanishes, the function $V_0\in\mathscr{V}_0$ does not go to zero fast enough
when $y\to+\infty$ to ensure the convergence of the
coefficients~\eqref{eq:coeffs}. To handle that case we use the lemma~7.2
of Ref.~\cite{BD86}. Following the same notation, we define by analytic
continuation the $B$-dependent function
\begin{equation}
R_{B}(r,s) = \frac{1}{2r_0^B}\int_0^r \ud x \,x^{B-3} \int_1^{+\infty}
\ud y \,V_0(y) \,F[s-x(y-1)] \,. \label{eq:RB}
\end{equation}
We can then write the leading term of the asymptotic expansion of the solution at
infinity, for any $V_0\in\mathscr{V}_0$, as
\begin{equation}
\mathop{\Psi}_{4,0}\!\!{}_{L=0} = \frac{1}{r}
\,\mathop{\mathrm{FP}}_{B=0} \int_{-\infty}^{t-r} \ud s
\,R_B\left(\frac{t-r-s}{2}, s\right) +
\mathcal{O}\left(\frac{1}{r^{2-\epsilon}}\right)\,. \label{eq:lemma}
\end{equation}
Inserting~\eqref{eq:RB} into~\eqref{eq:lemma} we get
\begin{equation}
\mathop{\Psi}_{4,0}\!\!{}_{L=0} = \frac{1}{2r}
\mathop{\mathrm{FP}}_{B=0} \int_{-\infty}^{t-r} \ud s
\int_0^{\frac{t-r-s}{2}} \ud x \,\frac{x^{B-3}}{r_0^B}\int_1^{+\infty}
\ud y \,V_0(y)\,F[s-x(y-1)] +
\mathcal{O}\left(\frac{1}{r^{2-\epsilon}}\right)\,.
\end{equation}
After the convenient change of variable $s\rightarrow z = \frac{t-r-s}{2}$, we
can integrate explicitly over $z$. Furthermore, we integrate three times by
part the remaining integral over $x$ so as to make the pole $\propto 1/B$
appear. Those operations result in
\begin{align}
\mathop{\Psi}_{4,0}\!\!{}_{L=0} &=
\mathop{\mathrm{FP}}_{B=0}\biggl\{\frac{1}{2B(B-1)(B-2)}\left(\frac{r}{r_0}
\right)^B \!\!\int_{1}^{+\infty} \ud s \,F^{(2)}(t-rs) (s-1)^B
\int_1^{+\infty} \ud y\,\frac{V_0(y)}{(y+1)^{B-2}}\biggr\} \nonumber
\\ &+
\mathcal{O}\left(\frac{1}{r^{2-\epsilon}}\right)\,. \label{eq:lemmanoconverge}
\end{align}
The above equation enables us to integrate all the terms that are not covered
by the previous formulas. Notice that the integral
\begin{equation}\label{eq:I0}
I_0(B) = \int_1^{+\infty} \ud y\,\frac{V_0(y)}{(y+1)^{B-2}}
\end{equation}
in~\eqref{eq:lemmanoconverge} diverges when $B=0$, since the function $V_0(y)$
only behaves like $y^{-1}$ when $y\to+\infty$. However, this divergence is
``protected'' by the analytic continuation in $B$ and it is even possible to
perform a Laurent expansion of $I_0(B)$ as $B$ goes to zero. To this aim, let
us consider the expansion of $V_0(y)$ in powers of the variable $y+1$ at
infinity. For the actual source we are interested in, it turns out that
\begin{equation} \label{eq:expansion_V0}
V_0(y) = \frac{V_{-1}}{y+1} + \frac{V_{-2}}{(y+1)^2} + \frac{V_{-3}}{(y+1)^3}
+ \frac{V^\text{log}_{-3} \ln (y+1)}{(y+1)^3} + \delta V_{-4}(y) \, ,
\end{equation}
where $V_{-1}$, $V_{-2}$, $V_{-3}$ and $V^\text{log}_{-3}$ are
numerical constants, whereas the function $\delta V_{-4}(y)$ behaves
like some power of $\ln (y-1)$ near $y=1$ and is $o(1/y^3)$ near $y
\to +\infty$. Thus, 
we have at first order in $B$:
\begin{equation} \label{eq:expansion_I0}
\int_1^{+\infty} \ud y\,\frac{\delta V_{-4}(y)}{(y+1)^{B-2}} = \int_1^{+\infty}
\ud y\, (y+1)^2\delta V_{-4}(y) \left[1-B \ln (y+1)\right]+\mathcal{O}\left(B^2\right)\, .
\end{equation}
Substituting to $V_0(y)$ its expansion~\eqref{eq:expansion_V0} in
Eq.~\eqref{eq:I0}, we find
\begin{align}
I_0(B) & = 
\frac{2^{2-B}V_{-1}}{B-2} + \frac{2^{1-B}V_{-2}}{B-1} + \frac{2^{-B}V_{-3}}{B}
+ 2^{-B} V_{-3}^\text{log} \left(\frac{\ln 2}{B} + \frac{1}{B^2} \right)
\nonumber \\ & + \int_1^{+\infty}
\ud y\, (y+1)^2\delta V_{-4}(y) \left[1-B \ln
  (y+1)\right]+\mathcal{O}\left(B^2\right) \, .
\end{align}
Each time that we have to apply Eq.~\eqref{eq:lemmanoconverge}, we use the
above truncated expression for the integral $I_0(B)$, which is readily
expanded up to the first order in $B$, and take the finite part when $B=0$ as
defined in~\eqref{eq:lemmanoconverge}.

\subsection{Integrating the instantaneous logarithmic terms}
\label{sec:log}

Finally, the quartic source also contains terms that are instantaneous, and
thus simpler than the previous hereditary terms, but involve the logarithm of
$r$. These instantaneous logarithmic terms are not covered by the
solution~\eqref{eq:solgen}. 
The problem amounts to finding an explicit representation of
\begin{align}
\mathop{\chi}_{k}\!{}_L &= \mathop{\mathrm{FP}}_{B=0} \,\Box^{-1}_R
\biggl[ \hat{n}_L \left( \frac{r}{r_0}\right)^B \ln \left(
  \frac{r}{r_0}\right) r^{-k}\,F(t - r) \biggr] \nonumber\\ &=
\mathop{\mathrm{FP}}_{B=0} \,\frac{\ud}{\ud B}\biggl\{\Box^{-1}_R
\biggl[ \hat{n}_L \left( \frac{r}{r_0}\right)^B r^{-k}\,F(t - r)
  \biggr]\biggr\}\,.\label{eq:logterm}
\end{align}
Notice that for all those terms the scale $r_0$ entering the instantaneous
logarithms $\ln(r/r_0)$ is the same as the one of our MPM algorithm.

\subsubsection{Case $k=2$}

According to Eq.~(A.2) of Ref.~\cite{B98quad} we have (for any
$B\in\mathbb{C}$):
\begin{equation}
\Box^{-1}_R \biggl[ \hat{n}_L \left( \frac{r}{r_0}\right)^B
  r^{-2}\,F(t - r) \biggr] = \frac{1}{K_\ell(B)} \int_{r}^{+\infty}
\ud s \,F(t-s) \,\hat{\partial}_L \left[\frac{(s-r)^{B+\ell} -
    (s+r)^{B+\ell}}{r} \right]\,,\label{eq:instkequal2}
\end{equation}
with $K_\ell(B) = 2(2r_0)^B B(B-1)\cdots(B-\ell)$; the operator
$\hat{\partial}_L$ denotes a STF product of $\ell$ spatial derivatives
($L=i_1\cdots i_\ell$). We inject~\eqref{eq:instkequal2}
into~\eqref{eq:logterm}, apply the differentiation with respect to $B$,
perform the Laurent expansion when $B\to 0$, and look for the finite part
coefficient. This leads to\footnote{In this derivation we use the fact that,
  for any integer such that $0\leqslant i\leqslant 2\ell$,~\cite{BD86}
$$\hat{\partial}_L \left[\frac{(s-r)^{i} - (s+r)^{i}}{r} \right] =
  0\,.$$}
\begin{align}
\mathop{\chi}_{2}\!{}_L &= \frac{(-)^\ell}{4\ell!} \int_{r}^{+\infty}
\ud s \,F(t-s) \,\hat{\partial}_L
\Biggl[\frac{(s-r)^\ell\left(\ln\bigl(\frac{s-r}{2r_0}\bigr)+H_\ell\right)^2
    -
    (s+r)^\ell\left(\ln\bigl(\frac{s+r}{2r_0}\bigr)+H_\ell\right)^2}{r}
  \Biggr]\,, \label{eq:logk2res}
\end{align}
where $H_\ell=\sum_{j=1}^\ell \frac{1}{j}$ is the $\ell$-th harmonic number.
An alternative, simpler representation of the right-hand side of
Eq.~\eqref{eq:logk2res} involving the Legendre function reads
\begin{align}
\mathop{\chi}_{2}\!{}_L &= - \frac{\hat{n}_L}{2 r} \int_{r}^{+\infty}
\ud s \,F(t-s) \,Q_\ell\Bigl(\frac{s}{r}\Bigr)
\biggl[\ln\left(\frac{s^2-r^2}{4r_0^2}\right) + 2H_\ell
  \biggr]\,. \label{eq:logk2resalt}
\end{align}
To prove it, we have verified that the above function satisfies the
requested d'Alembertian equation and has the same leading behaviour at
infinity as the expression~\eqref{eq:logk2res} of
$\mathop{\chi}_{2}\!{}_L$. As a result, the $1/r$ coefficient when $r
\to +\infty$ with $t-r$ constant can be computed either
from~\eqref{eq:logk2res} using the formulas (A.35) in~\cite{BD86}, or
more directly from~\eqref{eq:logk2resalt}, by inserting the
expansion~\eqref{eq:Qelly1} of the Legendre function $Q_\ell(y)$ when
$y\to 1^+$. We get
\begin{align}
\mathop{\chi}_{2}\!{}_L &= \frac{\hat{n}_L}{4r} \int_{0}^{+\infty} \ud
\tau \,F(t-r-\tau)
\biggl[\biggl(\ln\left(\frac{\tau}{2r_0}\right)+2H_\ell\biggr)^2 -
  \ln^2\left(\frac{r}{r_0}\right) \biggr] +
\mathcal{O}\left(\frac{1}{r^{2-\epsilon}}\right)\,. \label{eq:logk2exp}
\end{align}

\subsubsection{Case $3 \leqslant k \leqslant \ell+2$}

According to Eq.~(A9) of~\cite{B98quad} we have in this case:
\begin{align}
\Box^{-1}_R \biggl[ \hat{n}_L \left( \frac{r}{r_0}\right)^B
  r^{-k}\,F(t - r) \biggr] &= \left(\frac{r}{r_0}\right)^B
\,\sum_{i=0}^{k-3} \alpha_i(B) \,\hat{n}_L
\frac{F^{(i)}(t-r)}{r^{k-i-2}}\nonumber \\ &+ \beta(B) \,\Box^{-1}_R
\biggl[ \hat{n}_L \left( \frac{r}{r_0}\right)^B r^{-2}\,F^{(k-2)}(t -
  r) \biggr]\,.\label{eq:instkgeq3}
\end{align}
The $B$-dependent coefficients $\alpha_i(B)$ and $\beta(B)$ are given by
Eqs.~(A.10) of~\cite{B98quad}. However, for $3 \leqslant k \leqslant l+2$,
$\alpha_i(B)$ does not have any pole when $B\to 0$, and the expansion of
$\beta(B)$ starts at the first order in $B$, i.e., $\beta(B) = B +
\mathcal{O}(B^2)$. As the retarded integral of a source term whose radial
dependence is $r^{-2}$ (with any power of the logarithm of $r$) does not have
any pole either, we find
\begin{align}
\mathop{\chi}_{k}\!{}_L &= \frac{1}{r}\biggl[\alpha_{k-3}(0) \ln
  \left(\frac{r}{r_0} \right) + \alpha'_{k-3}(0) \biggr]
\hat{n}_L\,F^{(k-3)}(t-r) \nonumber \\ & + \beta'(0)
\mathop{\mathrm{FP}}_{B=0}\,\Box^{-1}_R \biggl[ \hat{n}_L \left(
  \frac{r}{r_0}\right)^B r^{-2}\,F^{(k-2)}(t - r) \biggr] +
\mathcal{O}\left(\frac{1}{r^{2-\epsilon}}\right)\,,\label{eq:logkgeq3}
\end{align}
where $\alpha'_{k-3}(0)$ and $\beta'(0)$ denote the $B$-derivative, evaluated
at $B=0$, of the coefficients $\alpha_{k-3}(B)$ and $\beta(B)$ displayed
explicitly in Ref.~\cite{B98quad}. For completeness, let us point out that
\begin{subequations}\label{eq:values1}
\begin{align}
\alpha_{k-3}(0) &= -
\frac{2^{k-3}(k-3)!(\ell-k+2)!}{(k-3-\ell)!(k-2+\ell)!}\,,\\ \alpha'_{k-3}(0)
&= \alpha_{k-3}(0) \Bigl[H_{k+\ell-2} - H_{k-3} -2H_\ell +
  H_{\ell-k+2}\Bigr]\,,
\end{align}\end{subequations}
with $\beta(0) = 0$ and $\beta'(0) = 2 \alpha_{k-3}(0)$. The
equation~\eqref{eq:logkgeq3} is sufficient for our purposes as we can compute
the last term thanks to the identity~\eqref{eq:instkequal2}.

\subsubsection{Case $k \geqslant \ell+3$}

In that case, Eq.~\eqref{eq:instkgeq3} is still valid, but $\alpha_{k-3}(B)$
has now a simple pole while $\beta(B)$ has no polar part. Let us then write
$\alpha_{k-3}(B)=a_{-1}B^{-1}+a_0+a_1 B + \mathcal{O}(B^2)$, so that
$\alpha'_{k-3}(B)=-a_{-1}B^{-2}+a_1+ \mathcal{O}(B)$; similarly,
$\beta(B)=b_0+b_1 B + \mathcal{O}(B^2)$ and $\beta'(B)=b_1+ \mathcal{O}(B)$.
When computing the finite part of~\eqref{eq:instkequal2} we are allowed to
commute the finite part operation with the evaluation of $\beta(B=0)=b_0$ and
$\beta'(B=0)=b_1$ since the retarded integral of a source term $\propto
r^{-2}$ is convergent. The solution $\mathop{\chi}_{k}\!{}_L$ may then be put
in the form
\begin{align}
\mathop{\chi}_{k}\!{}_L &= \frac{1}{r} \left[\frac{a_{-1}}{2} \ln^2
  \left(\frac{r}{r_0} \right) + a_0 \ln \left(\frac{r}{r_0} \right) +
  a_1\right] \hat{n}_L\,F^{(k-3)}(t-r) \nonumber \\ & + b_1
\mathop{\mathrm{FP}}_{B=0}\,\Box^{-1}_R \biggl[ \hat{n}_L \left(
  \frac{r}{r_0}\right)^B r^{-2}\,F^{(k-2)}(t - r) \biggr] \nonumber
\\ & + b_0 \mathop{\mathrm{FP}}_{B=0}\,\Box^{-1}_R \biggl[ \hat{n}_L
  \left( \frac{r}{r_0}\right)^B \ln \left(\frac{r}{r_0}\right)
  r^{-2}\,F^{(k-2)}(t - r) \biggr] +
\mathcal{O}\left(\frac{1}{r^{2-\epsilon}}\right)\,,
\label{eq:logkgeq32}
\end{align}
where
\begin{subequations}\label{eq:values2}
\begin{align}
a_{-1} &= \frac{(-)^{k+\ell}2^{k-3}(k-3)!}{(k-3-\ell)!(k-2+\ell)!}\,,
\\ a_0 &= a_{-1}\Bigl[H_{k-3-\ell} - H_{k-3} -2H_\ell +
  H_{k-2+\ell}\Bigr]\,,\\ a_1 &= \frac{a_0^2}{2a_{-1}} +
\frac{a_{-1}}{2}\Bigl[H_{k-3-\ell,2} - H_{k-3,2} + H_{k-2+\ell,2}\Bigr]\,,
\end{align}\end{subequations}
together with $b_0 = 2 a_{-1}$ and $b_1=2a_0$; here $H_{p,2}=\sum_{j=1}^{p}
\frac{1}{j^2}$ denotes the second harmonic number. The
formula~\eqref{eq:logkgeq32} is also sufficient for our purposes, as the
asymptotic form of the last two terms can be computed with the help of
Eqs.~\eqref{eq:instkequal2}--\eqref{eq:logk2res}.

\section{Quartic metric and radiative quadrupole moment} 
\label{sec:quartic}

Based on the formulas developed in Sec.~\ref{sec:explicit}, we have
implemented the MPM algorithm, as summarized by
Eqs.~\eqref{eq:umunu}--\eqref{eq:hmunu}, to compute the tails-of-tails cubic
metric $h_{M^2\times M_{ij}}$ for any $r$ greater than the radius of the
source. In particular, we have recovered the $1/r$ asymptotic
behaviour~\eqref{eq:asympt}, when $r \to +\infty$ with $t-r=$ const, obtained
in~\cite{B98tail}. With $h_{M^2\times M_{ij}}$ in hands, using
Eqs.~\eqref{eq:uquartic}--\eqref{eq:hmunuquartic}, we computed the quartic
source term $\Lambda_{M^3\times M_{ij}}$, checked that its divergence is
identically zero, and integrated it at leading order when $r \to +\infty$.
Within this stage, we have extensively employed the formulas developed in
Secs.~\ref{sec:asympt}--\ref{sec:log}.\footnote{For all these calculations we
  make intensive use of the algebraic computing software \textit{Mathematica}
  with the tensor package \textit{xAct}~\cite{xtensor}.}

Notice that once we have the quartic source $\Lambda_{M^3\times M_{ij}}$, we
could use the material from the previous section to determine the
comprehensive quartic metric $h_{M^3\times M_{ij}}$, but that would be a very
long process, since the function $V_m$ can be quite complex. Now, it is not
simply given by a Legendre function $Q_m$, as was the case with the cubic
metric $h_{M^2\times M_{ij}}$, but is typically a product of combinations of
Legendre functions with polynomials and algebraic rational fractions (such
that $V_m\in\mathscr{V}_m$, of course). As we are interested in the radiative
quadrupole moment detected at ``future null infinity'', we content ourselves
with the $1/r$ part of the quartic metric. This allows us to resort to the
far-zone version of the MPM algorithm defined in the Appendix~B
of~\cite{B98tail}.

Finally, our complete result for the leading $1/r$ term (actually made of
$1/r$, $\ln r/r$ and $\ln^2r/r$ terms) of the quartic metric in harmonic
coordinates reads
\begin{subequations}\label{eq:hM3Mij}
\begin{align}
h^{00}_{M^3\times M_{ij}} &= \frac{M^3\hat{n}_{ab}}{r}
\int_{0}^{+\infty} \ud\tau\, M^{(6)}_{ab} \left\{ - \frac{8}{3}
\ln^3\left(\frac{\tau}{2 r} \right) + \frac{148}{21} \ln^2
\left(\frac{\tau}{2 r}\right) + \frac{232}{21} \ln
\left(\frac{r}{r_0}\right )\ln \left(\frac{\tau}{2 r}\right) \right.
\nonumber \\ & \left.\quad\quad + \frac{1016}{2205}
\ln\left(\frac{\tau}{2r}\right) +
\frac{104}{15}\ln\left(\frac{r}{r_0}\right) + \frac{16489}{1575} -
\frac{232 \pi^2}{63} \right\} +
\mathcal{O}\left(\frac{1}{r^{2-\epsilon}}\right)\,, \\
h^{0i}_{M^3\times M_{ij}} &= \frac{M^3\hat{n}_{abi}}{r}
\int_{0}^{+\infty} \ud\tau\, M^{(6)}_{ab} \left\{ - \frac{26}{35}
\ln^2 \left(\frac{\tau}{2 r}\right) - \frac{8}{105} \ln
\left(\frac{\tau}{2r}\right) \ln \left(\frac{r}{r_0}\right)
\right. \nonumber \\ & \left.\quad\quad - \frac{6658}{11025}
\ln\left(\frac{\tau}{2r}\right)+ \frac{178}{315}
\ln\left(\frac{r}{r_0}\right) -\frac{59287}{33075} + \frac{8
  \pi^2}{315} \right\} \nonumber \\ &+ \frac{M^3\hat{n}_a}{r}
\int_{0}^{+\infty} \ud\tau\, M^{(6)}_{ai} \left\{ - \frac{8}{3} \ln^3
\left(\frac{\tau}{2 r}\right) + \frac{562}{75} \ln^2
\left(\frac{\tau}{2 r} \right) + \frac{832}{75} \ln
\left(\frac{\tau}{2r} \right) \ln \left(\frac{r}{r_0} \right)
\right. \nonumber \\ &\left.\quad\quad + \frac{926}{1125}
\ln\left(\frac{\tau}{2r}\right) + \frac{1154}{175}
\ln\left(\frac{r}{r_0}\right) + \frac{212134}{18375} - \frac{832
  \pi^2}{225}\right\} +
\mathcal{O}\left(\frac{1}{r^{2-\epsilon}}\right)\,, \\
h^{ij}_{M^3\times M_{ij}} &= \frac{M^3\hat{n}_{abij}}{r}
\int_{0}^{+\infty} \ud\tau\, M^{(6)}_{ij} \left\{ -
\ln^2\left(\frac{\tau}{2 r}\right) - \frac{4}{5} \ln
\left(\frac{\tau}{2 r}\right) + \frac{107}{105} \ln
\left(\frac{r}{r_0}\right) - \frac{30868}{11025}\right\}
\nonumber\\ &+ 2\frac{M^3\hat{n}_{a(j}}{r} \int_{0}^{+\infty}
\ud\tau\, M^{(6)}_{i)a} \left\{ \frac{234}{35} \ln^2
\left(\frac{\tau}{2 r}\right) + \frac{104}{35}
\ln\left(\frac{\tau}{2r}\right) \ln\left(\frac{r}{r_0}\right)
\right. \nonumber\\ &\left.\quad\quad + \frac{58694}{3675}
\ln\left(\frac{\tau}{2 r}\right) - \frac{598}{735} \ln
\left(\frac{r}{r_0}\right) + \frac{1487812}{77175} - \frac{104
  \pi^2}{105} \right\} \nonumber\\ &+ \frac{M^3\hat{n}_{ab}
  \delta_{ij}}{r} \int_{0}^{+\infty} \ud\tau\, M^{(6)}_{ab} \left\{ -
\frac{48}{7} \ln^2 \left( \frac{\tau}{2 r}\right) - \frac{64}{21} \ln
\left(\frac{\tau}{2 r} \right) \ln \left(\frac{r}{r_0} \right)
\right. \nonumber \\ &\left.\quad\quad - \frac{7108}{441} \ln
\left(\frac{\tau}{2 r} \right) + \frac{1756}{2205} \ln
\left(\frac{r}{r_0} \right) -\frac{4508029}{231525} +
\frac{64\pi^2}{63} \right\} \nonumber \\ &+ \frac{M^3}{r}
\int_{0}^{+\infty} \ud\tau\, M^{(6)}_{ij} \left\{ - \frac{8}{3} \ln^3
\left(\frac{\tau}{2 r} \right) + \frac{16}{3} \ln^2
\left(\frac{\tau}{2 r} \right) + \frac{152}{15} \ln
\left(\frac{\tau}{2 r} \right) \ln \left(\frac{r}{r_0} \right)
\right. \nonumber \\ & \left.\quad\quad - \frac{2332}{525} \ln
\left(\frac{\tau}{2 r} \right) + \frac{3608}{525} \ln
\left(\frac{r}{r_0} \right) + \frac{286408}{55125} - \frac{152
  \pi^2}{45}\right \} +
\mathcal{O}\left(\frac{1}{r^{2-\epsilon}}\right)\,.
\end{align}
\end{subequations}
The quadrupole moments inside the integrals are evaluated at time $t - r -
\tau$. Note that, at this stage, the logarithms involve both the radial
distance $r$ to the source and the constant $r_0$ coming from the MPM
algorithm. We shall now extract the relevant physical information from the
above metric as viewed at future null infinity, in the form of the so-called
radiative quadrupole moment $U_{ij}$~\cite{Th80} --- not to be confused of
course with the source type quadrupole moment $M_{ij}$.

So far, we have performed all our computations in harmonic coordinates
$x^\mu$. However, this choice of coordinates has the well-known disadvantage
that the coordinate cones $t-r$ (where $r = \vert x^i \vert$) deviate by
powers of the logarithm of $r$ from the true space-time characteristics or
light cones. As a result, the $1/r$ expansion of the metric (as $r \to
+\infty$ with $t-r=$ const) involves powers of logarithms. We get rid of them
by going to radiative coordinates $X^\mu$ for which the associated coordinate
cones $T -R$ (where $R = \vert X^i \vert$) become asymptotically tangent to
the true light cones at future null infinity. As in previous
works~\cite{B98tail}, this is achieved by applying the coordinate
transformation $X^\mu = x^\mu + \xi^\mu(x)$, where $\xi^\mu$ is defined by
\begin{subequations}\label{eq:defxi}
\begin{align}
\xi^0 &= -2 M \ln\left(\frac{r}{b_0} \right)\,,\\\xi^i &= 0\,,
\end{align}\end{subequations}
with $b_0$ denoting an arbitrary scale that is \textit{a priori} different
from the scale $r_0$. Let us show that this simple coordinate change is
sufficient to remove all the log-terms from our quartic
metric~\eqref{eq:hM3Mij}, so that, in radiative coordinates $X^\mu$, it is
straightforward to define the radiative quadrupole moment $U_{ij}$. Keeping
only the $1/R$ terms and consistently taking into account all the $M^3\times
M_{ij}$ interactions, one can check that the metric in radiative coordinates
$H_{M^3\times M_{ij}}$ differs from the metric $h_{M^3\times M_{ij}}$ in
harmonic-coordinates by
\begin{align}
H^{\mu\nu}_{M^3\times M_{ij}} &= h^{\mu\nu}_{M^3\times M_{ij}} -
\xi^\lambda \partial_\lambda h^{\mu\nu}_{M^2\times M_{ij}} +
\frac{1}{2} \xi^\lambda \xi^\sigma \partial^2_{\lambda\sigma}
h^{\mu\nu}_{M\times M_{ij}} - \frac{1}{6}\xi^\lambda
\xi^\sigma\xi^\rho \partial^3_{\lambda\sigma\rho} h^{\mu\nu}_{M_{ij}}
+ \mathcal{O}\left(\frac{1}{R^2}\right)\,,\label{eq:radiativecoord}
\end{align}
where both sides are evaluated at the same dummy coordinate point, say
$X^\mu$. Injecting in this relation the results found for $h_{M^3\times
  M_{ij}}$, $h_{M^2\times M_{ij}}$, $h_{M\times M_{ij}}$ and $h_{M_{ij}}$,
recalled in Sec.~\ref{sec:recalls}, with $\xi^\mu$ given by~\eqref{eq:defxi},
we indeed observe that all the logarithms of $R$ vanish. More precisely, we
obtain
\begin{subequations}\label{eq:HM3Mij}
\begin{align}
H^{00}_{M^3\times M_{ij}} &= \frac{M^3\hat{N}_{ab}}{R}
\int_{0}^{+\infty} \ud\tau\, M^{(6)}_{ab} \left\{ - \frac{8}{3}
\ln^3\left(\frac{\tau}{2 b_0} \right) - 4 \ln^2 \left(\frac{\tau}{2
  b_0}\right) + \frac{232}{21} \ln \left(\frac{\tau}{2 b_0}\right )\ln
\left(\frac{\tau}{2 r_0}\right) \right.  \nonumber \\ &
\left.\quad\quad  - \frac{14272}{2205}\ln\left(\frac{\tau}{2b_0}\right) + 
\frac{104}{15} \ln\left(\frac{\tau}{2r_0}\right) +
\frac{16489}{1575} - \frac{232 \pi^2}{63}\right\} +
\mathcal{O}\left(\frac{1}{R^2}\right)\,, \\
H^{0i}_{M^3\times M_{ij}} &= \frac{M^3\hat{N}_{abi}}{R}
\int_{0}^{+\infty} \ud\tau\, M^{(6)}_{ab} \left\{ - \frac{2}{3} \ln^2
\left(\frac{\tau}{2 b_0}\right) - \frac{8}{105} \ln
\left(\frac{\tau}{2b_0}\right) \ln \left(\frac{\tau}{2 r_0}\right)
\right. \nonumber \\ & \left.\quad\quad - \frac{1432}{1225}
\ln\left(\frac{\tau}{2b_0}\right) + \frac{178}{315}
\ln\left(\frac{\tau}{2 r_0}\right) -\frac{59287}{33075} + \frac{8
  \pi^2}{315} \right\} \nonumber \\ &+ \frac{M^3\hat{N}_a}{R}
\int_{0}^{+\infty} \ud\tau\, M^{(6)}_{ai} \left\{ - \frac{8}{3} \ln^3
\left(\frac{\tau}{2 b_0}\right) - \frac{18}{5} \ln^2
\left(\frac{\tau}{2 b_0} \right) + \frac{832}{75} \ln
\left(\frac{\tau}{2 b_0} \right) \ln \left(\frac{\tau}{2 r_0} \right)
\right. \nonumber \\ &\left.\quad\quad -\frac{45448}{7875}
\ln\left(\frac{\tau}{2 b _0}\right) + \frac{1154}{175}
\ln\left(\frac{\tau}{2 r _0}\right) + \frac{212134}{18375} - \frac{832
  \pi^2}{225}\right\} + \mathcal{O}\left(\frac{1}{R^2}\right)\,, \\
H^{ij}_{M^3\times M_{ij}} &= \frac{M^3\hat{N}_{abij}}{R}
\int_{0}^{+\infty} \ud\tau\, M^{(6)}_{ab} \left\{ -
\ln^2\left(\frac{\tau}{2 b_0}\right) - \frac{191}{105} \ln
\left(\frac{\tau}{2 b_0}\right) + \frac{107}{105} \ln
\left(\frac{\tau}{2 r_0}\right) -\frac{30868}{11025}\right\} \nonumber
\\ &+ 2\frac{M^3\hat{N}_{a(j}}{R} \int_{0}^{+\infty} \ud\tau\,
M^{(6)}_{i)a} \left\{ \frac{26}{7} \ln^2 \left(\frac{\tau}{2 b_0}\right) +
\frac{104}{35} \ln\left(\frac{\tau}{2 b_0}\right)
\ln\left(\frac{\tau}{2 r_0}\right)
\right. \nonumber\\ &\left.\quad\quad + \frac{8812}{525}
\ln\left(\frac{\tau}{2 b_0}\right) - \frac{598}{735} \ln
\left(\frac{\tau}{2 r_0}\right) + \frac{1487812}{77175} - \frac{104
  \pi^2}{105}\right\} \nonumber\\ &+ \frac{M^3\hat{N}_{ab}
  \delta_{ij}}{R} \int_{0}^{+\infty} \ud\tau\, M^{(6)}_{ab} \left\{ -
\frac{80}{21} \ln^2 \left( \frac{\tau}{2 b_0}\right) - \frac{64}{21}
\ln \left(\frac{\tau}{2 b_0} \right) \ln \left(\frac{\tau}{2 r_0}
\right) \right. \nonumber \\ &\left.\quad\quad - \frac{592}{35} \ln
\left(\frac{\tau}{2 b_0} \right) + \frac{1756}{2205} \ln
\left(\frac{\tau}{2 r_0} \right) -\frac{4508029}{231525} +
\frac{64\pi^2}{63} \right\} \nonumber \\ &+ \frac{M^3}{R}
\int_{0}^{+\infty} \ud\tau\, M^{(6)}_{ij} \left\{ - \frac{8}{3} \ln^3
\left(\frac{\tau}{2 b_0} \right) - \frac{24}{5} \ln^2
\left(\frac{\tau}{2 b_0} \right) + \frac{152}{15} \ln
\left(\frac{\tau}{2 b_0} \right) \ln \left(\frac{\tau}{2 r_0} \right)
\right. \nonumber \\ & \left.\quad\quad - \frac{396}{35} \ln
\left(\frac{\tau}{2 b_0} \right) + \frac{3608}{525} \ln
\left(\frac{\tau}{2 r_0} \right) +\frac{286408}{55125} - \frac{152
  \pi^2}{45}\right \} + \mathcal{O}\left(\frac{1}{R^2}\right)\,,
\end{align}
\end{subequations}
where the quadrupole moments are evaluated at time $T_R-\tau$ in the past,
with $T_R=T-R$ denoting the retarded time in radiative coordinates.

By definition, the radiative mass and current multipole moments $U_L(T_R)$ and
$V_L(T_R)$ are then the multipolar coefficients that parameterize the
transverse-tracefree (TT) projection of the spatial metric in radiative
coordinates, at retarded radiative time $T_R$, i.e.,
\begin{equation}
H_{ij}^\text{TT} = - \frac{4}{R} \,\mathcal{P}_{ijkl}
\sum_{\ell=2}^{+\infty} \frac{1}{\ell!} \left\{ N_{L-2}
\,U_{klL-2}(T_R) - \frac{2\ell}{\ell+1} N_{aL-2}
\,\varepsilon_{ab(k}\,V_{l)bL-2}(T_R) \right\} +
\mathcal{O}\left(\frac{1}{R^2}\right)\,.
\end{equation}
The TT projection operator is given by $\mathcal{P}_{ijkl} =
\mathcal{P}_{ik}\mathcal{P}_{jl} -
\frac{1}{2}\mathcal{P}_{ij}\mathcal{P}_{kl}$ where
$\mathcal{P}_{ij}=\delta_{ij} - N_i N_j$ represents the projector onto the
plane transverse to the unit direction $N_i=X_i/R$ from the source to the
observer. The associated total energy flux $\mathcal{F}=(\ud E/\ud
T_R)^\text{GW}$ reads~\cite{Th80}
\begin{equation}\label{eq:FluxF}
\mathcal{F} = \sum^{+\infty}_{\ell=2} \biggl[
  \frac{(\ell+1)(\ell+2)}{(\ell-1)\ell \ell!(2\ell+1)!!}
  \bigl(U^{(1)}_L\bigr)^2 + \frac{4\ell
    (\ell+2)}{(\ell-1)(\ell+1)!(2\ell+1)!!}  \bigl(V^{(1)}_L\bigr)^2
  \biggr]\,.
\end{equation}

For the contribution of the tails-of-tails-of-tails part of the radiative
metric~\eqref{eq:HM3Mij} to the radiative quadrupole moment $U_{ij}$, we get
\begin{align}
\delta U_{ij}(T_R) &= M^3 \int_{0}^{+\infty} \ud \tau
\,M^{(6)}_{ij}(T_R - \tau) \left[\frac{4}{3} \ln^3 \left(\frac{\tau}{2
    b_0} \right) + \frac{11}{3} \ln^2 \left(\frac{\tau}{2 b_0} \right)
  + \frac{124627}{11025} \ln\left(\frac{\tau}{2b_0}\right)
  \right. \nonumber \\ &\left. \qquad\qquad -\frac{428}{105}
  \ln\left(\frac{\tau}{2b_0}\right) \ln \left(\frac{\tau}{2r_0}\right)
  - \frac{1177}{315} \ln\left(\frac{\tau}{2r_0}\right) +
  \frac{129268}{33075} + \frac{428}{315}\pi^2\right]
\,.\label{eq:dUij}
\end{align}
Adding the known quadratic tails and cubic tails-of-tails~\cite{B98tail}, we
obtain the radiative mass quadrupole moment, complete with respect to such
tail interactions up to the quartic level:
\begin{align}
U_{ij}(T_R) &= M^{(2)}_{ij}(T_R) + \frac{GM}{c^3} \int^{+\infty}_0 \ud
\tau \,M^{(4)}_{ij} (T_R-\tau) \left[ 2\ln \left( \frac{c\tau}{2b_0}
  \right) + \frac{11}{6} \right] \nonumber\\&\qquad+
\frac{G^2M^2}{c^6} \int^{+\infty}_0 \ud \tau \,M^{(5)}_{ij}(T_R-\tau)
\left[ 2\ln^2 \left( \frac{c\tau}{2b_0} \right) + \frac{11}{3} \ln
  \left( \frac{c\tau}{2b_0} \right)
  \right.\nonumber\\&\left. \qquad\qquad\qquad\qquad - \frac{214}{105}
  \ln \left( \frac{c\tau}{2r_0} \right) + \frac{124627}{22050} \right]
\nonumber\\&\qquad+ \frac{G^3M^3}{c^9} \int_{0}^{+\infty} \ud \tau
\,M^{(6)}_{ij}(T_R - \tau) \left[\frac{4}{3} \ln^3
  \left(\frac{c\tau}{2 b_0} \right) + \frac{11}{3} \ln^2
  \left(\frac{c\tau}{2 b_0} \right) \right. \nonumber
  \\ &\left. \qquad\qquad\qquad\qquad + \frac{124627}{11025}
  \ln\left(\frac{c\tau}{2b_0}\right) -\frac{428}{105}
  \ln\left(\frac{c\tau}{2b_0}\right) \ln
  \left(\frac{c\tau}{2r_0}\right) \right. \nonumber
  \\ &\left. \qquad\qquad\qquad\qquad - \frac{1177}{315}
  \ln\left(\frac{c\tau}{2r_0}\right) + \frac{129268}{33075} +
  \frac{428}{315}\pi^2\right] +
\mathcal{O}\left(\frac{1}{c^{12}}\right)\,.\label{eq:radquad}
\end{align}
We have restored the powers of $G$ and $c$ to show that the
tails-of-tails-of-tails represent a 4.5PN effect in the waveform. They
correspond to the most difficult interaction between multipole moments to be
computed up to the 4.5PN level. However, there are several other types of
interactions that are easier to control and have not been included here. These
are for example the non-linear memory
integrals~\cite{B90,Chr91,WW91,Th92,BD92,B98quad,F09,F11} starting at the
2.5PN order, but also many instantaneous terms, notably at the 4PN order. All
these contributions will be systematically investigated in future work.

We have checked that the coefficient of the maximal power of the logarithm in
the quartic tails of~\eqref{eq:dUij}--\eqref{eq:radquad} (namely, the cubic
logarithm with coefficient $4/3$) agrees with the expectation for the dominant
iterated infra-red type logarithms, usually factorized out in tail-induced
resummed waveforms~\cite{DIN09,FBI15} (see e.g., Sec.~(3.1) in~\cite{FBI15}).

For future reference let us also recall the radiative mass octupole and
current quadrupole radiative moments up to the cubic tails~\cite{FBI15}
\begin{subequations}\label{eq:radoctcurr}
\begin{align}
U_{ijk}(T_R) &= M^{(3)}_{ijk}(T_R) + \frac{GM}{c^3} \int^{+\infty}_0
\ud \tau \,M^{(5)}_{ijk} (T_R-\tau) \left[ 2\ln \left(
  \frac{c\tau}{2b_0} \right) + \frac{97}{30} \right]
\nonumber\\&\qquad+ \frac{G^2M^2}{c^6} \int^{+\infty}_0 \ud \tau
\,M^{(6)}_{ijk}(T_R-\tau) \left[ 2\ln^2 \left( \frac{c\tau}{2b_0}
  \right) + \frac{97}{15} \ln \left( \frac{c\tau}{2b_0} \right)
  \right.\nonumber\\&\left. \qquad\qquad\qquad\qquad - \frac{26}{21}
  \ln \left( \frac{c\tau}{2r_0} \right) + \frac{13283}{4410}
  \right] + \mathcal{O}\left(\frac{1}{c^9}\right)\,,\\
V_{ij}(T_R) &= S^{(2)}_{ij}(T_R) + \frac{GM}{c^3} \int^{+\infty}_0 \ud
\tau \,S^{(4)}_{ij} (T_R-\tau) \left[ 2\ln \left( \frac{c\tau}{2b_0}
  \right) + \frac{7}{3} \right] \nonumber\\&\qquad+ \frac{G^2M^2}{c^6}
\int^{+\infty}_0 \ud \tau \,S^{(5)}_{ij}(T_R-\tau) \left[ 2\ln^2
  \left( \frac{c\tau}{2b_0} \right) + \frac{14}{3} \ln \left(
  \frac{c\tau}{2b_0} \right)
  \right.\nonumber\\&\left. \qquad\qquad\qquad\qquad - \frac{214}{105}
  \ln \left( \frac{c\tau}{2r_0} \right) - \frac{26254}{11025}
  \right] + \mathcal{O}\left(\frac{1}{c^9}\right)\,.
\end{align}\end{subequations}

\section{Energy flux of compact binaries on circular orbits}
\label{sec:flux}

In this section we derive, based on the quartic radiative mass quadrupole
moment~\eqref{eq:radquad}, the complete 4.5PN coefficient of the
gravitational-wave energy flux~\eqref{eq:FluxF}, in the case of binary systems
of non-spinning compact objects moving on circular orbits. We thus extend the
circular energy flux known at the 3.5PN
order~\cite{BDIWW95,B98tail,BFIJ02,BDEI04} by including the 4.5PN coefficient,
while the determination of the 4PN coefficient is left to future work. The
test mass limit of our new 4.5PN coefficient turns out to be in perfect agreement
with the prediction from black-hole perturbation
theory~\cite{TTS96,Fuj14PN,Fuj22PN}.

The reason why we are able to control the 4.5PN order without knowing
the complete 4PN field (since the source moments are known only up to
the 3.5PN order~\cite{BFIS08,FMBI12,FBI15}) is the fact that for
half-integral PN orders, i.e., $\frac{n}{2}$PN orders where $n$ is an
\textit{odd} integer, any instantaneous or ``non-hereditary'' term is
zero in the energy flux for circular orbits. This can be shown by a
simple dimensional argument (see the discussion in Sec.~II of
Ref.~\cite{BFW14a}). Notice that memory effects do not arise at
half-integral PN orders, since the time derivative acting on the
radiative moments in the flux equation~\eqref{eq:FluxF} turns them
into instantaneous quantities. Hence, at the 4.5PN order, only truly
``hereditary'' tail integrals do contribute to the circular energy
flux. It is therefore sufficient to control the occurrence of such
hereditary integrals, i.e., of non-linear tail interactions between
multipole moments. For that purpose, it is very useful to apply some
``selection rules'' that permit one to determine all the possible
multipole interactions occurring at a given PN
order~\cite{BFIS08,FMBI12,FBI15} (see in particular Sec.~III
of~\cite{FBI15}).

According to those selection rules, in order to control the 4.5PN order for
circular orbits, we need only the contributions of (i) quadratic multipole
tails, of the form $M\times M_L$ or $M\times S_L$, with
$2\leqslant\ell\leqslant 5$ for mass moments $M_L$ and
$2\leqslant\ell\leqslant 4$ for current moments $S_L$, (ii) the quartic
quadrupole tails-of-tails-of-tails $M^3\times M_{ij}$ (with $\ell=2$ in this
case), and (iii) the double product between the quadratic quadrupole tails
$M\times M_{ij}$ and the cubic quadrupole tails-of-tails $M^2\times M_{ij}$.

The cubic tails-of-tails $M^2\times M_{ij}$ by themselves contribute to the
circular energy flux, starting at the 3PN order~\cite{B98tail}. The
4PN order correction will be given in Eq.~\eqref{eq:cubtail}. Moreover,
from the point (iii) above, we see that the cubic tails-of-tails also contribute
at the 4.5PN order through their interactions with the quadrupole tails
$M\times M_{ij}$.

The computation of quadratic tails for circular compact binaries is
classic and will not be detailed. Suffice it to say that, at the 4.5PN
order, we need the mass quadrupole moment at 3PN order, since 4.5PN
means 3PN beyond the dominant quadrupole tail at the 1.5PN order. The 3PN
quadrupole moment for circular orbits reads (see e.g.,
Ref.~\cite{BFIS08})\footnote{Here $x^i$ and $v^i$ denote the orbital
  separation and relative velocity of the two particles (and the
  angular brackets refer to the STF projection). The mass parameters are
  the total mass $m=m_1+m_2$ and the symmetric mass ratio
  $\nu=\frac{m_1m_2}{(m_1+m_2)^2}$. The harmonic-coordinates PN
  parameter is $\gamma=\frac{G m}{r c^2}$, where $r=\vert x^i\vert$ represents
  the radial harmonic-coordinates separation. The quasi-invariant PN
  parameter is $x=(\frac{G m\omega}{c^3})^{2/3}$ where $\omega$ stands for the
  orbital frequency, related to $r$ by Eqs.~\eqref{eq:omega}. A scalar
  such as the circular energy flux is \textit{quasi-invariant}
  when expressed in terms of $x$, in the sense that it stays invariant
  under the class of coordinate transformations that are
  asymptotically Minkowskian at infinity.}
\begin{equation}
M_{ij} = m \,\nu \left(A \, x^{\langle ij \rangle}+B \,
\frac{r^2}{c^2}v^{\langle ij \rangle} +
\frac{48}{7}\frac{G^2m^2\nu}{c^5 r} x^{\langle i}v^{j \rangle}\right)
+ \mathcal{O}\left(\frac{1}{c^7}\right)\,.
  \label{eq:Mij}
\end{equation}
In order to control the tails at the 4.5PN order we may ignore the above 2.5PN
dissipative term since it contributes only at the 4PN order. Notice that
the ``canonical'' quadrupole moment $M_{ij}$ agrees for circular orbits, up to
the 3PN order, with the alternative definition of the ``source'' quadrupole
moment $I_{ij}$~\cite{BFIS08}. The two coefficients $A$ and $B$ are given by
the following expansion series in the PN parameter $\gamma$:
\begin{subequations}\label{eq:AB}
  \begin{align}
    A &= 1 + \gamma \left(-\frac{1}{42} - \frac{13}{14}\nu \right) +
    \gamma^2 \left(-\frac{461}{1512} - \frac{18395}{1512}\nu -
    \frac{241}{1512} \nu^2\right) \\ & + \gamma^3 \left(
    \frac{395899}{13200} - \frac{428}{105} \ln \left(
    \frac{r}{r_0}\right) + \left[ \frac{3304319}{166320} -
      \frac{44}{3} \ln \left(\frac{r}{{r'}_0} \right) \right] \nu +
    \frac{162539}{16632} \nu^2 + \frac{2351}{33264}\nu^3
    \right)\,,\nonumber\\
B &= \frac{11}{21} - \frac{11}{7} \nu + \gamma \left( \frac{1607}{378}
- \frac{1681}{378} \nu + \frac{229}{378} \nu^2 \right) \nonumber\\ & +
\gamma^2 \left( - \frac{357761}{19800} + \frac{428}{105} \ln \left(
\frac{r}{r_0} \right) - \frac{92339}{5544} \nu + \frac{35759}{924}
\nu^2 + \frac{457}{5544} \nu^3 \right)\,.
  \end{align}
\end{subequations}
Notice the two scales entering the logarithmic terms at the 3PN order: one is
the length scale $r_0$ coming from the MPM algorithm (see
Sec.~\ref{sec:recalls}), while the other scale $r'_0$ is the logarithmic
barycenter of two gauge constants $r'_1$ and $r'_2$ which appear in the 3PN
equations of motion in harmonic coordinates~\cite{BFeom}, i.e., $m\ln r'_0 =
m_1\ln r'_1 + m_2\ln r'_2$. The latter constant $r'_0$ thus parameterizes the
relation between the orbital frequency $\omega$ for circular orbits and the
separation $r$ at the 3PN order in harmonic coordinates, namely
\begin{subequations}\label{eq:omega}
\begin{align}
  \omega^2 &= \frac{Gm}{r^3} \bigg\{ 1+\bigl(-3+\nu\bigr) \gamma +
  \left( 6 + \frac{41}{4}\nu + \nu^2 \right) \gamma^2 \\ & \qquad +
  \left( -10 + \left[- \frac{75707}{840} + \frac{41}{64} \pi^2 + 22
    \ln \left( \frac{r}{r'_0}\right) \right]\nu + \frac{19}{2}\nu^2 +
  \nu^3 \right) \gamma^3 + \mathcal{O}\left(\frac{1}{c^8}\right)
  \biggr\} \,,\nonumber\\ \gamma &= x
  \biggl\{1+\left(1-{\nu\over3}\right)x + \left(1-{65\over 12} \nu
  \right) x^2 \\ & \qquad + \left( 1 + \left[-{2203\over 2520}-
    {41\over 192}\pi^2 - {22\over 3}\ln\left(r\over{r'}_0\right)
    \right] \nu +{229\over 36}\nu^2+ {1\over 81}\nu^3\right)x^3 +
  \mathcal{O}\left(\frac{1}{c^8}\right) \biggr\}\,.\nonumber
\end{align}\end{subequations}
Here, we do not consider the  2.5PN radiation reaction term in the
equations of motion, since it generates some contribution at the 4PN
order but not at the 4.5PN order. To summarize, there are three arbitrary length
scales in the problem: $r_0$, $r'_0$, as well as $b_0$ which originates from our
choice of radiative type coordinate system through
Eqs.~\eqref{eq:defxi}. It is non trivial to check that these three
scales cancel out in the final gauge invariant expression of the
energy flux for circular orbits.

Another important step of the calculation is the reduction of the tail
integrals to circular orbits. As usual, those integrals are to be computed
proceeding as if the worldlines in the integrands obeyed the current circular
dynamics, which amounts to neglecting the evolution in the past by radiation
reaction. The influence of the past evolution would be to correct the dominant
1.5PN tail integral by a 2.5PN radiation reaction term, and would thus be of
order 4PN, but not 4.5PN. From Eq.~\eqref{eq:radquad}, we see that, at the
4.5PN order, we also need some integration formulas involving up to three
powers of logarithms. Those are~\cite{GR}
\begin{subequations}\label{eq:intlog}
  \begin{align}
     \int^{+\infty}_0 \!\!\ud\tau\,\ln
     \left(\frac{\tau}{\tau_0}\right)\, e^{-\ui\Omega\tau} &=
     \frac{\ui}{\Omega} \left(\ln\bigl(\Omega\tau_0\bigr)
     +\gamma_\text{E}+\ui\frac{\pi}{2}\right)\,, \\ \int^{+\infty}_0
     \!\!\ud\tau\,\ln^2\left(\frac{\tau}{\tau_0}\right) \,
     e^{-\ui\Omega\tau} &= - \frac{\ui}{\Omega} \left[
       \left(\ln\bigl(\Omega\tau_0\bigr)
       +\gamma_\text{E}+\ui\frac{\pi}{2}\right)^2 +
       \frac{\pi^2}{6}\right]\,,\\ \int^{+\infty}_0
     \!\!\ud\tau\,\ln^3\left(\frac{\tau}{\tau_0}\right) \,
     e^{-\ui\Omega\tau} &= \frac{\ui}{\Omega} \left[
       \left(\ln\bigl(\Omega\tau_0\bigr)
       +\gamma_\text{E}+\ui\frac{\pi}{2}\right)^3 +
       \frac{\pi^2}{2}\left(\ln\bigl(\Omega\tau_0\bigr)
       +\gamma_\text{E}+\ui\frac{\pi}{2}\right)
       +2\zeta(3)\right]\,.\label{eq:intlog3}
\end{align}
\end{subequations}
Here, $\Omega$ denotes a multiple of the orbital frequency
$\omega$. The constant $\tau_0$ is arbitrary, and related either to
$r_0$, $r'_0$ or $b_0$.  We denote by $\gamma_\text{E} \simeq 0.577$
the Euler constant, whereas $\zeta(3) \simeq 1.202$ is the Ap{\'e}ry
constant ($\zeta$ being the usual notation for the Riemann zeta
function).

Let us decompose the tail contributions to the energy flux of circular
binaries up to the 4.5PN order --- as defined by~\eqref{eq:FluxF} in
terms of the radiative moments --- into those generated by quadratic,
cubic and quartic tails,
\begin{equation}
\mathcal{F}_\text{tail} = \mathcal{F}_\text{quadratic} +
\mathcal{F}_\text{cubic} + \mathcal{F}_\text{quartic} +
\mathcal{O}\left(G^5\right)\,,
\end{equation}
where the remainder contains the neglected MPM approximations [see
Eq.~\eqref{eq:MPM}]. The quadratic tails correspond to multipole interactions
$M\times M_L$ and $M\times S_L$ (see e.g., Eqs.~(3.6)--(3.7) in~\cite{FBI15}).
In particular, we need the full 3PN precision for the mass quadrupole moment,
as in Eqs.~\eqref{eq:Mij}--\eqref{eq:AB}. The higher order moments require
some lower PN precision. Their explicit expressions can be found in Ref.~\cite{BFIS08}.
It is also worthy to note that the quadratic tails contribute only to half-integral PN
approximations. The result in terms of the PN parameter $\gamma=G m/(r
  c^2)$ up to the 4.5PN order reads (factorizing out the Newtonian flux as usual)
\begin{align}
  \mathcal{F}_\text{quadratic} &= \frac{32c^5}{5G}\nu^2 \gamma^5
  \biggl\{ 4\pi \gamma^{3/2} \nonumber \\ & \qquad +
  \left(-\frac{25663}{672}-\frac{125}{8}\nu\right)\pi \gamma^{5/2} +
  \left(\frac{90205}{576} + \frac{505747}{1512}\nu +
  \frac{12809}{756}\nu^2\right)\pi \gamma^{7/2} \nonumber \\ & \qquad
  + \left(\frac{9997778801}{106444800} -
  \frac{6848}{105}\ln\left(\frac{r}{r_0}\right) +\left[ -
    \frac{8058312817}{2661120} + \frac{287}{32}\pi^2 +
    \frac{572}{3}\ln\left(\frac{r}{r'_0}\right)\right]\nu
  \right. \nonumber \\ & \qquad \qquad \left. -
  \frac{12433367}{13824}\nu^2 - \frac{1026257}{266112}\nu^3 \right)\pi
  \gamma^{9/2} + \mathcal{O}\left(\frac{1}{c^{11}}\right)\biggr\}\,.
  \label{eq:quadtail}
\end{align}
In contrast to the quadratic tails, the cubic tails-of-tails contribute to
integral PN approximations, starting at the 3PN order~\cite{B98tail}. At the
next 4PN order they involve the contribution of the mass quadrupole moment (to
be computed with 1PN precision), as well as that of the mass octupole and
current quadrupole moments given by Eqs.~\eqref{eq:radoctcurr}. Furthermore,
at the same level of the cubic tails, we must include in the flux the square
of the quadratic tails. Those various contributions have all been computed. For their
sum, we obtain, extending Eq.~(5.9) of~\cite{B98tail},
\begin{align}
  \mathcal{F}_\text{cubic} &= \frac{32c^5}{5G}\nu^2 \gamma^5
  \biggl\{\left( - \frac{116761}{3675} + \frac{16}{3}\pi^2 -
  \frac{1712}{105}\gamma_\text{E} - \frac{856}{105} \ln(16\gamma) +
  \frac{1712}{105} \ln\left(\frac{r}{r_0}\right) \right)\gamma^3
  \nonumber\\ & \qquad + \left( \frac{12484937}{30870} -
  \frac{4040}{63}\pi^2 + \frac{86456}{441}\gamma_\text{E} +
  \frac{43228}{441} \ln(16\gamma) - \frac{86456}{441}
  \ln\left(\frac{r}{r_0}\right) \right. \nonumber\\ & \qquad\qquad +
  (1-4\nu)\left[\frac{670000393}{7408800} + \frac{445}{42}\pi^2 -
    \frac{56731}{4410}\gamma_\text{E} - \frac{56731}{8820}
    \ln(16\gamma) \right. \nonumber\\ & \qquad\qquad \left.\left. +
    \frac{133771}{4410} \ln2 - \frac{47385}{1568} \ln3 +
    \frac{56731}{4410} \ln\left(\frac{r}{r_0}\right)
    \right]\right)\gamma^4 +
  \mathcal{O}\left(\frac{1}{c^{10}}\right)\biggr\}\,.
  \label{eq:cubtail}
\end{align}
The constant $b_0$ disappears, as expected. However,
$\mathcal{F}_\text{cubic}$ still contains $r_0$. The point is that these cubic
tails at the 3PN and 4PN orders are not the only contributions to the full
coefficients, since the flux also contains many instantaneous (non-tails)
terms that depend on the source multipole moments, and notably the 4PN
quadrupole. After the moments have been replaced by their explicit
expressions, those terms should cancel out the remaining constants $r_0$
in~\eqref{eq:cubtail}. Thus, since the 4PN instantaneous contributions are not
known, we shall ignore henceforth the 4PN coefficient in the flux except for
the partial result~\eqref{eq:cubtail}.

In addition, there are other tail contributions at the 4PN order (but
not at the 4.5PN order) that we have not yet taken into account. We
can mention for instance the coupling between the dominant 1.5PN tail
term and the 2.5PN non-linear memory effect, which is therefore
expected to contribute at the 4PN order. Moreover, there exists a
2.5PN effect corresponding to the past evolution of the binary source
due to radiation reaction. It should affect the computation of the
1.5PN tail integral at the 4PN order. We also recall the non-local 4PN
tail term entering the equations of motion~\cite{BBBFM16a,BBBFM16b},
which will have to be included when performing the order reduction of
accelerations coming from the time derivatives of the Newtonian
quadrupole moment. All these contributions will have to be
systematically included in future work.

Let us next focus on the computation of the quartic-order tails in the
flux.  One contribution is directly due to the quartic tail term at
the 4.5PN order in the radiative mass quadrupole
moment~\eqref{eq:radquad}. However, there is another contribution
coming from a double product between the quadratic quadrupole tail at
the 1.5PN order and the cubic quadrupole tail-of-tail at the 3PN order
--- recall that the energy flux contains the square of the time
derivative of Eq.~\eqref{eq:radquad}. It turns out that important
cancellations occur between these two terms, notably all the
logarithms squared and cubed disappear, leaving only a term linear in
the logarithm. The constant $b_0$ cancels out as expected, but a
dependence of $r_0$ is left out at this stage. In the end, we find
that
\begin{align}
  \mathcal{F}_\text{quartic} &= \frac{32c^5}{5G}\nu^2 \gamma^5
  \biggl\{\left( - \frac{467044}{3675} - \frac{3424}{105}
  \ln(16\gamma) + \frac{6848}{105} \ln\left(\frac{r}{r_0}\right) -
  \frac{6848}{105}\gamma_\text{E} \right)\pi \gamma^{9/2}
  \nonumber\\ &\qquad\qquad\qquad +
  \mathcal{O}\left(\frac{1}{c^{11}}\right)\biggr\}\,.
  \label{eq:quartail}
\end{align}

Finally we are in a position to control the half-integral PN approximations
(or so-called ``odd'' PN terms) in the energy flux for circular orbits, as
they are entirely due to tail integrals. The
``odd'' part of the flux in this case is
\begin{align}
  \mathcal{F}{\Big|}_\text{odd} &=
  \mathcal{F}_\text{tail}{\Big|}_\text{odd} =
  \Bigl(\mathcal{F}_\text{quadratic} +
  \mathcal{F}_\text{quartic}\Bigr){\Big|}_\text{odd} +
  \mathcal{O}\left(G^5\right)\,.
  \label{eq:Fodddef}
\end{align}
We do not include the cubic tail part~\eqref{eq:cubtail} since it is
``even'' in the PN sense. Therefore, we need only to sum up
Eqs.~\eqref{eq:quadtail} and~\eqref{eq:quartail}. We gladly
discover that the scale $r_0$ cancels out from the sum, thereby
obtaining
\begin{align}
  \mathcal{F}{\Big|}_\text{odd} &= \frac{32c^5}{5G}\nu^2 \gamma^5
  \biggl\{ 4\pi \gamma^{3/2} \nonumber\\ & \qquad \qquad \quad +
  \left(-\frac{25663}{672}-\frac{125}{8}\nu\right)\pi \gamma^{5/2} +
  \left(\frac{90205}{576} + \frac{505747}{1512}\nu +
  \frac{12809}{756}\nu^2\right)\pi \gamma^{7/2} \nonumber \\ & \qquad
  \qquad \quad + \left( - \frac{24709653481}{745113600} -
  \frac{6848}{105}\gamma_\text{E} -
  \frac{3424}{105}\ln\left(16\gamma\right) \right.\nonumber \\ &
  \qquad \qquad\qquad\qquad \quad \left.+\left[ -
    \frac{8058312817}{2661120} + \frac{287}{32}\pi^2 +
    \frac{572}{3}\ln\left(\frac{r}{r'_0}\right)\right]\nu
  \right.\nonumber \\ & \qquad \qquad\qquad\qquad \quad \left.-
  \frac{12433367}{13824}\nu^2 - \frac{1026257}{266112}\nu^3 \right)\pi
  \gamma^{9/2} + \mathcal{O}\left(\frac{1}{c^{11}}\right)\biggr\}\,.
  \label{eq:fluxtotalgam}
\end{align}
Still there remains a dependence on the scale $r'_0$ coming from the
equations of motion, but that is merely due to our use of the
harmonic-coordinates PN parameter $\gamma$. Eliminating $\gamma$ in
favor of the quasi-invariant frequency-related PN parameter
$x=(G m\omega/c^3)^{2/3}$ with the help of Eqs.~\eqref{eq:omega} yields
then our final result:
\begin{align}
  \mathcal{F}{\Big|}_\text{odd} &= \frac{32c^5}{5G}\nu^2 x^5 \biggl\{
  4\pi x^{3/2} \nonumber \\ & \qquad \qquad \quad +
  \left(-\frac{8191}{672}-\frac{583}{24}\nu\right)\pi x^{5/2} +
  \left(-\frac{16285}{504} + \frac{214745}{1728}\nu +
  \frac{193385}{3024}\nu^2\right)\pi x^{7/2} \nonumber \\ & \qquad
  \qquad \quad + \left( \frac{265978667519}{745113600} -
  \frac{6848}{105}\gamma_\text{E} -
  \frac{3424}{105}\ln\left(16x\right) +\left[ \frac{2062241}{22176} +
    \frac{41}{12}\pi^2\right]\nu \right.\nonumber \\ & \qquad
  \qquad\qquad\qquad \quad \left.- \frac{133112905}{290304}\nu^2 -
  \frac{3719141}{38016}\nu^3 \right)\pi x^{9/2} +
  \mathcal{O}\left(\frac{1}{c^{11}}\right)\biggr\}\,.
  \label{eq:Foddx}
\end{align}
We insist that the latter odd part of the flux, although it has been
computed only from tail contributions, represents the full PN-odd part of the
complete flux, in the case of circular orbits. Thus, the
coefficients can be compared with those derived from black-hole
perturbation theory in the small mass ratio limit $\nu\to
0$. Black-hole perturbations have been expanded for this problem at
the 1.5PN order~\cite{P93a}, then extended up to the 5.5PN order
in~\cite{TNaka94,Sasa94,TSasa94,TTS96}, and more recently, using the
method~\cite{MST96a,MST96b,MT97}, up to extremely high PN
orders~\cite{Fuj14PN,Fuj22PN}. Our new 4.5PN result in
Eq.~\eqref{eq:Foddx} perfectly reproduces the latter works in the
limit where $\nu\to 0$ (see Eq.~(3.1) in~\cite{TTS96}).

\section{Conclusion} 
\label{sec:concl}

This paper is a contribution to our current program to provide 4.5PN
accurate gravitational waveforms (together with the orbital phasing)
generated by the inspiral of compact binary systems without spins. A
first part of this program, concerning the 4PN accurate equations of
motion, has already been completed~\cite{BBBFM16a,BBBFM16b}.  Here, we
solve one of the main difficulties regarding the 4.5PN wave field,
namely the computation of \textit{quartic} non-linearities associated
with high order tail effects called ``tails-of-tails-of-tails''. These
terms correspond to the interaction between three mass monopoles $M$
and the quadrupole moment $M_{ij}$. They contribute at the 4.5PN order
to the asymptotic waveform and gravitational energy flux (beyond the
Einstein quadrupole formula).

Our calculation is based on the multipolar-post-Minkowskian (MPM)
algorithm for solving the Einstein field equations in the exterior
region of a general isolated source~\cite{BD86,B98quad,B98tail}. We
developed new mathematical formulas to express the retarded solutions
of d'Alembertian equations sourced by certain non-linear tail (or
``hereditary'') integrals. Such formulas are necessary in practice for
implementing the non-linear iterations leading to the quartic metric,
and allow us to express our results in essentially analytic closed
form. Those formulas involve a machinery of Legendre polynomials and
associated Legendre functions.

We have thoroughly computed the cubic-order tails-of-tails at any distance
from the source (while only the leading asymptotic terms were previously
known~\cite{B98tail}), and plugged the latter piece of the gravitational field
into the source term in order to perform the next iteration of the metric at
the quartic level. From that source term, we derived, resorting again to
formulas generalized from previous works, the tails-of-tails-of-tails at the
leading order in the inverse distance to the source. Our main result is
encapsulated in the radiative mass quadrupole moment~\eqref{eq:radquad}, which
describes the quadrupolar gravitational waves at infinity, up to the level of
the quartic interaction $M^3\times M_{ij}$ which contributes at the 4.5PN
order in the asymptotic waveform.

As an application, we computed the total energy flux emitted by
compact binary systems in the case of circular orbits. After extending
the classic calculation of quadratic tails up to the 4.5PN order, we
obtained the new contributions due to the quartic
tails-of-tails-of-tails. We also extended the computation of cubic
tails-of-tails at the 4PN order. This led us to the complete 4.5PN
coefficient in the energy flux of compact binaries on circular orbits,
as given by Eq.~\eqref{eq:Foddx}. The energy flux represents an
essential theoretical input for gravitational-wave data analysis,
since it drives the orbital phase evolution~\cite{3mn,CF94}. Notice,
however, that the 4PN coefficient is not yet known, except in
the test mass limit. Its computation is left for future work. Finally,
the test-particle limit of our 4.5PN expression (when $\nu\to 0$) is
in perfect agreement with the result found by means of black-hole
perturbation methods applied to the two body
problem~\cite{TTS96,Fuj14PN,Fuj22PN}.


\appendix

\section{Proofs of some mathematical formulas}

\subsection{Formula~\eqref{eq:kgeq2}}
\label{app:proof1}

We want to prove that, for any $\ell \geqslant k-2$, $m \geqslant
k-2$, and for any $V_m \in \mathscr{V}_m$,
\begin{align}
\mathop{\Psi}_{k,m}\!\!{}_L &= - \hat{n}_L \int_{1}^{+\infty} \ud s
\,F^{(k-2)}(t-rs) \nonumber\\&\qquad\qquad\times\biggl[ Q_\ell (s)
  \int_{1}^s\ud y \,V_m^{(-k+2)}(y) P_\ell(y) + P_\ell(s)
  \int_{s}^{+\infty} \ud y \,V_m^{(-k+2)}(y)
  Q_\ell(y)\biggr]\,,\label{eq:A1app}
\end{align}
where $V_m^{(-k+2)}(y)$ is the $(k-2)$-th anti-derivative of $V_m(y)$ defined
by Eq.~\eqref{eq:primVp}. We will proceed by induction over the integer $k$.
Let us thus assume that~\eqref{eq:A1app} is valid up to $k-1$ with $k
\geqslant 3$, and let us show that it is then valid for $k$. By definition, we have
\begin{equation}
\mathop{\Psi}_{k,m}\!\!{}_L = \mathop{\mathrm{FP}}_{B=0} \,\Box^{-1}_R
\biggl[ \hat{n}_L \left( \frac{r}{r_0}\right)^B r^{-k}
  \int_{1}^{+\infty} \ud y \,V_m(y) F(t - ry) \biggr]\,,
\end{equation}
which we can integrate by part, for $V_m \in \mathscr{V}_m$ and $m \geqslant
k-2$. We find
\begin{equation}\label{eq:intpart}
\mathop{\Psi}_{k,m}\!\!{}_L = \mathop{\mathrm{FP}}_{B=0}\Box^{-1}_R
\biggl[ \hat{n}_L \left( \frac{r}{r_0}\right)^B r^{-k+1}
  \int_{1}^{+\infty} \ud y \,V_m^{(-1)}(y) F^{(1)}(t - ry)\biggr]\,,
\end{equation}
where $V_m^{(-1)}(y)=\int_{1}^y \ud x \,V_m(x)$ in agreement
with~\eqref{eq:primVp}. As $m \geqslant 1$ (because $k \geqslant 3$),
$\alpha_m = \int_{1}^{+\infty} \ud y \,V_m(y)$ is a convergent integral.
Posing $\tilde{V}_{m-1}(y)=V_m^{(-1)}(y)-\alpha_m$, we rewrite~\eqref{eq:intpart} as
\begin{align} 
\mathop{\Psi}_{k,m}\!\!{}_L &= \alpha_m
\,\mathop{\mathrm{FP}}_{B=0}\Box^{-1}_R \biggl[ \hat{n}_L \left(
  \frac{r}{r_0}\right)^B r^{-k} F(t-r)\biggr]\nonumber \\ & +
\mathop{\mathrm{FP}}_{B=0}\Box^{-1}_R \biggl[ \hat{n}_L \left(
  \frac{r}{r_0}\right)^B r^{-k+1} \int_{1}^{+\infty} \ud y
  \,\tilde{V}_{m-1}(y) \,F^{(1)}(t - ry) \biggr] \,.
\end{align}
The point now is that $\tilde{V}_{m-1} \in \mathscr{V}_{m-1}$, so that we
can make use of Eq.~\eqref{eq:A1app} (which is our induction
hypothesis) to obtain
\begin{align}
\mathop{\Psi}_{k,m}\!\!{}_L &= \alpha_m
\,\mathop{\mathrm{FP}}_{B=0}\Box^{-1}_R \biggl[ \hat{n}_L \left(
  \frac{r}{r_0}\right)^B r^{-k} F(t-r)\biggr]\nonumber \\ & -
\hat{n}_L \int_{1}^{+\infty} \ud s \,F^{(k-2)}(t-rs)
\nonumber\\&\qquad\qquad\times\biggl[ Q_\ell (s) \int_{1}^s\ud y
  \,\tilde{V}_{m-1}^{(-k+3)}(y) P_\ell(y) + P_\ell(s)
  \int_{s}^{+\infty} \ud y \,\tilde{V}_{m-1}^{(-k+3)}(y)
  Q_\ell(y)\biggr]\,.\label{eq:rec}
\end{align}
The first term is instantaneous and, keeping in mind that $\ell
\geqslant k-2$, it may be integrated by means of the formula~(A.11)
of~\cite{B98quad}. Since we have
$\tilde{V}_{m-1}^{(-k+3)}(y)=V_m^{(-k+2)}(y)-\alpha_m\frac{(y-1)^{k-3}}{(k-3)!}$,
we obtain exactly the result ${}_{k,m}\Psi_L$ given
by~\eqref{eq:A1app} that we wanted to prove, with however the
following additional term
\begin{align}\label{eq:additional}
\delta\!\!\mathop{\Psi}_{k,m}\!\!{}_L &= \alpha_m \,\hat{n}_L
\int_{1}^{+\infty}\ud s \,F^{(k-2)}(t-rs)
\nonumber\\&\qquad\qquad\qquad\times\left[\mathcal{A}_{\ell}^{k-3}(s)
  - \frac{2^{k-3} (k-3)!  (\ell+2-k)!}{(\ell+k-2)!}  \sum_{j=0}^{k-3}
  \,\frac{(\ell+j)!\,(s-1)^j}{2^j (\ell-j)!(j!)^2} \right]\,,
\end{align}
in which we have introduced the following combination
\begin{equation}
\mathcal{A}_{\ell}^{k-3}(s) = Q_\ell(s) \int_{1}^s\ud y \,P_\ell(y)
\,\frac{(y-1)^{k-3}}{(k-3)!} + P_\ell(s) \int_{s}^{+\infty}\ud y
\,Q_\ell(y) \,\frac{(y-1)^{k-3}}{(k-3)!}\,.\label{eq:Aellp}
\end{equation}

We are now going to prove that the additional
term~\eqref{eq:additional} is actually zero, because the quantity in
the square brackets of~\eqref{eq:additional} is in fact identically
zero (for any $s \in ] 1, +\infty[$). To this end, we notice that
    the two integrals appearing in~\eqref{eq:Aellp} are of the same
    type, namely
\begin{equation}
\mathcal{I}_{\ell}^{p}(s) = \int_{a}^s\ud y \,f_\ell(y)
\,\frac{(y-1)^{p}}{p!}\,,
\end{equation}
where we have posed $p=k-3$ for simplicity sake. The lower boundary is $a=1$
or $a=+\infty$ according to the integral in~\eqref{eq:Aellp} we
are considering. The function $f_\ell(y)$ represents either the Legendre
polynomial $P_\ell(y)$ for $a=1$ or the Legendre function $Q_\ell(y)$ for
$a=+\infty$. In both cases, the integral is  well-defined. Using the
fact that $f_\ell(y)$ satisfies the usual Legendre differential
equation\footnote{Namely,
\begin{equation*}
\frac{\ud}{\ud y}\biggl[(1-y^2)\frac{\ud f_\ell(y)}{\ud
    y}\biggr]+\ell(\ell+1)f_\ell(y)=0\,.
\end{equation*}
We remind also the following properties of Legendre functions that are used in
our computation:
\begin{align}
&(1-y^2)\frac{\ud f_\ell(y)}{\ud y} = \ell\Bigl[f_{\ell-1}(y)-y
    f_\ell(y)\Bigr]\,,\nonumber\\ &P_{\ell}(y)Q_{\ell-1}(y) -
  P_{\ell-1}(y)Q_{\ell}(y) = \frac{1}{\ell}\,.\nonumber
\end{align}
}
and performing two integrations by parts, we obtain the following recursive
relation
\begin{equation}
\mathcal{I}_{\ell}^{p}(s) = \frac{1}{(\ell-p)(\ell+p+1)}
\biggl[\Bigl(f_\ell(s)\bigl[(\ell-p)s-p\bigr]-\ell
  f_{\ell-1}(s)\Bigr)\frac{(s-1)^{p}}{p!} + 2
  p\,\mathcal{I}_{\ell}^{p-1}(s)\biggr]\,.\label{eq:intrecI}
\end{equation}
It nicely translates, when inserted into~\eqref{eq:Aellp}, into the
simple recurrence equation
\begin{equation}
\mathcal{A}_{\ell}^{p}(s) =
\frac{1}{(\ell-p)(\ell+p+1)}\Bigl[\frac{(s-1)^{p}}{p!} + 2
  p\,\mathcal{A}_{\ell}^{p-1}(s)\Bigr]\,,\label{eq:intrecA}
\end{equation}
whose solution is straightforwardly found to be
\begin{equation}
\mathcal{A}_{\ell}^{p}(s) = \frac{2^{p} p!  (\ell-p-1)!}{(\ell+p+1)!}
\sum_{j=0}^{p} \,\frac{(\ell+j)!\,(s-1)^j}{2^j
  (\ell-j)!(j!)^2}\,.\label{eq:solAellp}
\end{equation}
To arrive at the latter expression, we need the readily checked relation
$A_{\ell}^{0}(s)=\frac{1}{\ell(\ell+1)}$, which plays the role of
normalization condition. The result~\eqref{eq:solAellp} shows that the
additional term~\eqref{eq:additional} is indeed zero, which completes our
proof of Eq.~\eqref{eq:A1app}.

\subsection{Formulas~\eqref{eq:k3} and~\eqref{eq:plusgen}}
\label{app:proof2}

We notice that, for the proof of the case where $\ell=0$, $k=3$ and
$m=2$, all the reasonings in the previous section~\ref{app:proof1}
remain valid for $\ell=0$ up to Eq.~\eqref{eq:rec}, which is actually
true as soon as $m\geqslant 1$. Now, the equation~(A.13)
of~\cite{B98quad} tells us that
\begin{equation}
\mathop{\mathrm{FP}}_{B=0}\Box^{-1}_R \biggl[ \left(
  \frac{r}{r_0}\right)^B r^{-3} F(t-r)\biggr] = -
\biggl[\ln\left(\frac{r}{r_0}\right)+1\biggr]\frac{F(t-r)}{r} + 2
\int_{1}^{+\infty} \ud s \,F^{(1)}(t-rs)Q_0(s)\,.
\label{eq:k3l0}
\end{equation}
Inserting~\eqref{eq:k3l0} into~\eqref{eq:rec} for $k=3$, we readily
recover the result~\eqref{eq:k3} in the text.

The generalization to the case where $\ell=0$, $k\geqslant 3$ and $m\geqslant
k-2$, given by the formula~\eqref{eq:plusgen} in the text, differs from the proof
we just presented for $k=3$, and goes as follows. First, using
the definition~(6.3) of Ref.~\cite{BD86}, we pose
\begin{equation}
R_B(r,s) = \frac{1}{2} \int_0^r \ud x \left(\frac{x}{r_0} \right)^B
x^{-k+1}\int_1^{+\infty} \ud y \,V_m(y) F[s -
  x(y-1)]\,. \label{eq:eqR}
\end{equation}
Then, by virtue of the theorem (6.1) of~\cite{BD88}, we can write
\begin{align}
\mathop{\Psi}_{k,m}\!\!{}_{L=0} &= \frac{1}{r}\,\mathrm{FP}_{B=0}
\int_{-\infty}^{t-r} \ud s \left[R_B\left(\frac{t-r-s}{2}, s\right) -
  R_B\left(\frac{t+r-s}{2}, s\right)\right]\,.
\end{align}
Let us call $A_1$ the first term in this expression, actually
a retarded homogeneous solution of the wave equation, and $A_2$ the
second term, which is made of a mixture of retarded and advanced
times. By construction, we have ${}_{k,m}\Psi_{L=0} = A_1 + A_2$.

We inject~\eqref{eq:eqR} into $A_1$ and define the new variable $u
= \frac{t-r-s}{2}$, thereby expressing $A_1$ in terms of the new set
of variables $(u,x,y)$. Next, after commuting the $x$ and $u$ integrals, we
explicitly integrate over $u$. Once this is done, we perform $k-1$
integrations by part with respect to $x$ and apply the finite part
procedure to get
\begin{equation} 
A_1 = \frac{(-)^k}{2(k-2)!} \frac{1}{r}\int_{0}^{+\infty}\ud x
\biggl[\ln \left(\frac{x}{r_0}\right) + H_{k-2}
  \biggr]\int_{1}^{+\infty} \ud y \,V_m(y) \,(y+1)^{k-1}
\,F^{(k-2)}\bigl[t-r-x(y+1)\bigr] \,. \label{eq:A1}
\end{equation}
The same treatment is applied to the second term $A_2$ but the
computation is longer, as some boundary terms arise. We
find that
\begin{align} 
A_2 &= - \frac{(-)^{k}}{2(k-2)!} \frac{1}{r}\int_{r}^{+\infty}\ud x
\biggl[\ln \left(\frac{x}{r_0}\right) + H_{k-2}
  \biggr]\!\!\int_{1}^{+\infty} \!\!\ud y \,V_m(y) \,(y+1)^{k-1}
\,F^{(k-2)}\bigl[t+r-x(y+1)\bigr] \nonumber\\ & -
\frac{(-)^{k}}{2(k-2)!} \frac{1}{r}\int_{0}^{r}\ud x \biggl[\ln
  \left(\frac{x}{r_0}\right) + H_{k-2} \biggr]\!\!\int_{1}^{+\infty}
\!\!\ud y \,V_m(y) \,(y-1)^{k-1}
\,F^{(k-2)}\bigl[t-r-x(y-1)\bigr]\nonumber\\ & +
\frac{(-)^{k}}{(k-2)!} \frac{1}{r} \biggl[\ln
  \left(\frac{r}{r_0}\right) + H_{k-2} \biggr]\int_{1}^{+\infty} \!\ud
y \,V_m(y) \,\varphi_{k-2}(y)\,F^{(k-3)}(t-r y)\nonumber\\ & -
\sum_{i=1}^{k-3} \frac{(-)^i}{(k-2) \cdots (k-2-i)}
\frac{1}{r^{k-i-1}} \int_{1}^{+\infty} \!\ud y\,V_m(y) \,\varphi_i(y)
\,F^{(i-1)}(t-ry)\,,\label{eq:A2}
\end{align}
where
$\varphi_{i}(y)=\frac{1}{2}[(y+1)^{i}-(y-1)^{i}]$. Then, we make the
change of variable $t-rs = t - r - x(y +1)$ in~\eqref{eq:A1}, as well as the
changes $t-rs = t + r - x(y +1)$ in the first line of~\eqref{eq:A2}
and $t-rs = t - r - x(y -1)$ in the second line
of~\eqref{eq:A2}. Finally, with these new variables, after
exchanging the integrations,  simplifications occur, yielding
the formula~\eqref{eq:plusgen} in the text.

\bibliography{ListeRef.bib}

\end{document}